\documentclass[journal,draftclsnofoot,onecolumn,12pt,twoside]{IEEEtranTCOM}
\normalsize

\usepackage{wrapfig}

\usepackage[caption=false,font=footnotesize]{subfig}
\usepackage{epsfig} 

\usepackage{amsmath,texdef2004} 
\usepackage{amssymb}  
\usepackage{float}
\usepackage{cases,setspace}
\usepackage{adjustbox}
\usepackage{graphicx}
\usepackage[lined,ruled,commentsnumbered]{algorithm2e}
\usepackage{multicol}
\newcommand{\SINR}{\text{SINR}}

\newcommand{\WIFI}[1]{R^{(W)}_{#1}}

\ifCLASSINFOpdf
\else
\fi

\usepackage{url}

\begin{document}
%
\title{Optimizing Networks for Internet Access Using Tethering}
%
%
%

\author{Vandana Mittal$^{*}$, Sanjit K. Kaul$^{**}$ and Sumit Roy$^{\dagger}$\\
	$^{*}$ECE Dept., University of Manitoba,\\	
	$^{**}$Wireless Systems Lab, IIIT-Delhi,\
	$^{\dagger}$University of Washington, Seattle, WA\\
	mittalv@myumanitoba.ca, skkaul@iiitd.ac.in, sroy@u.washington.edu}
\maketitle

\begin{abstract}
We investigate scenarios where Internet access to a user device (node) is available only via the cellular network. However, not every node may connect directly to it. Instead, some may use tethering to connect over WiFi to a node sharing its Internet connection. In effect, nodes split into hotspots and clients. Hotspots are nodes that connect directly to the cellular network and can provide Internet connectivity to other nodes to whom they are connected over WiFi. Clients connect to the cellular network only via hotspots. In this work, we consider the problem of determining the split of hotspots and clients, and the association between them, which maximizes the sum of the rates of all nodes, subject to the constraint that any node gets at least the rate it gets when all nodes are directly connected to the cellular network. Via tractable networks, we provide insights into the interplay between WiFi connectivity amongst nodes and rates of their links to the cellular tower, the splits that maximize sum rate, with provably optimal splits for a few cases. We propose a novel heuristic approach to split any network and provide a detailed exposition of gains available from tethering, via simulations. 
\end{abstract}


%
\IEEEpeerreviewmaketitle

\section{Introduction}
\label{chapter:introduction}

\begin{figure*}[t]
	\begin{center}
		\subfloat[]{\includegraphics[height=.22\columnwidth]{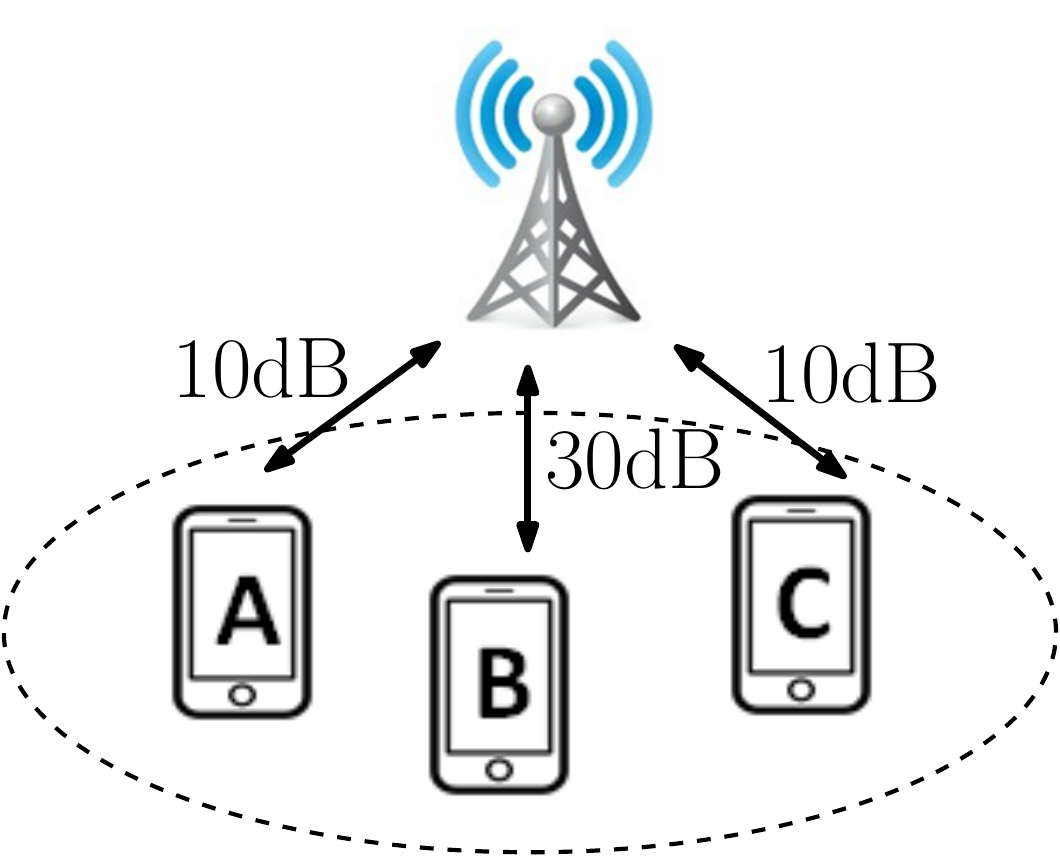}
			\label{fig:example1}%
		} \hspace{35pt}
		\renewcommand{\thesubfigure}{b}
		\subfloat[]{\includegraphics[height=.22\columnwidth]{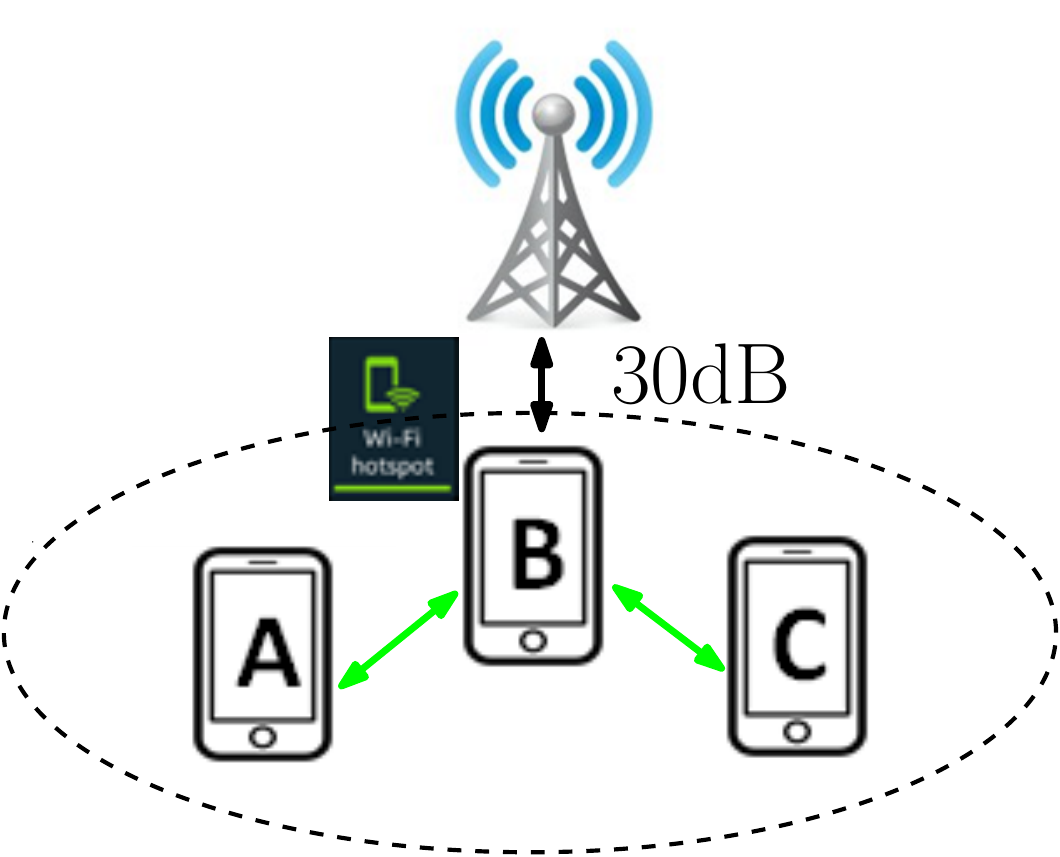}
			\label{fig:example1a}%
		} \hspace{35pt}
		\renewcommand{\thesubfigure}{c}
		\subfloat[]{\includegraphics[height=.22\columnwidth]{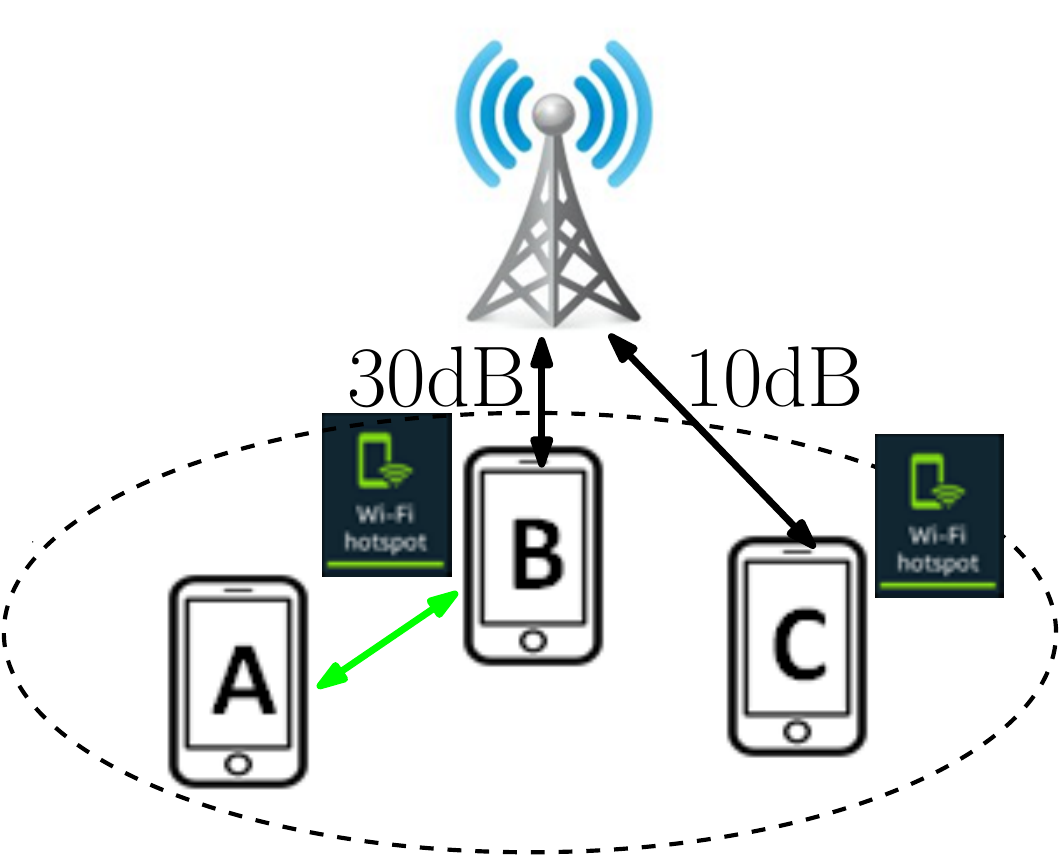}
			\label{fig:example2a}%
	    }		
		\caption{\small(a) Example \emph{baseline} network of three nodes $A$, $B$, and $C$. All nodes \emph{use} their links to the cell tower. The SINR of a link is mentioned besides it. Network (b) has $B$ as the hotspot and $A$ and $C$ as clients. Network (c) has $B$ and $C$ as hotspots and $A$ is a client of $B$. The green links are the WiFi links in use in the hotspot networks.}
		\label{fig:examples}
	\end{center}
		\vspace{-0.55in}
\end{figure*}

Mobile consumer devices (smartphones etc.) often come with WiFi and cellular (3G/4G) radios. A commonly available feature on such devices (nodes) is {\em mobile tethering}, whereby a group of nodes can be organized into a \emph{hotspot network} (see Figure~\ref{fig:example1a}), which consists of clients and hotspots. For connectivity to the Internet, the hotspots use their own links to the cell tower (cellular base station), whereas the clients use the hotspots' links to the cell tower. The clients connect to the hotspots over WiFi (green links in Figure~\ref{fig:example1a}). In this work, we leverage tethering to organize a network of nodes into hotspots and clients such that the \emph{sum rate} of the network is maximized. Our scenario is distinct from the traditional {\em data offloading} mechanisms in cellular access~\cite{qutqut2013mfw}~\cite{siris2013enhancing}. There a home or enterprise {\em indoor} user offloads their cellular data traffic to a WiFi access point (AP) in the vicinity; the WiFi AP uses it's wired backhaul (cable/fiber/ethernet) to connect to the Internet independently of the cellular access network.  Our work is aimed typically at {\em outdoor} scenarios, where Internet access is only via the (wireless) cellular network. 


%


Next we introduce the key aspects of our model, which we formalize in Section~\ref{chapter:OptimizationProblem}. We consider a network consisting of a single cell tower and one or more nodes. Each node has a link to the tower. We limit ourselves to when all links to the tower share the same channel. A node also has a WiFi link to every other node. Only hotspots \emph{use their link} (occupy the cellular resource for a non-zero time share) to the tower and clients \emph{use} their WiFi links to their hotspots. A hotspot and all clients connecting to it over WiFi are each allocated a rate on the hotspot's link to the tower\footnote{For ease of exposition, we don't distinguish between uplink and downlink of a node.}. Clearly, the sum of allocated rates cannot exceed the rate of the hotspot's link to the tower. Also, a rate allocated to a client must not exceed the rate of its WiFi link to the hotspot. We assume every node always has a packet to send and saturates any allocated rate.

We define a \emph{baseline} network (for example, Figure~\ref{fig:example1}) as one in which all nodes use their links to the tower, that is tethering is not used. The rate allocated to a node is therefore the rate of its link to the tower. Note that, with respect to \emph{baseline}, tethering reduces the number of nodes (\emph{load}) that use their links to the tower (see Figures~\ref{fig:example1a} and~\ref{fig:example2a}). It, however, increases the number of nodes, and corresponding links to their hotspots, that share the WiFi spectrum. Also, links differ in \emph{quality} (in this work, a known link signal-to-interference-and-noise-ratio (SINR)). To analyze tethering, we must therefore capture load and quality for both WiFi and cellular. 

This we achieve by setting the rate of a link to the product of its spectral efficiency and the share of time it is used. The spectral efficiency of a link is set to its Shannon rate and is an increasing function of quality. Without loss of generality, we assume that all hotspots get the same time share to the tower. Such a rate allocation model may be interpreted as a fair allocation of cellular resource amongst the hotspots, as has been suggested earlier~\cite{singh_offloading_2012}. Also, all clients connecting to a hotspot (over WiFi) get the same time share to the hotspot. As nodes always have data to send, this can be achieved via packet aggregation~\cite{Bing:2007:ETW:1564932}. Specifically, clients with WiFi links having larger spectral efficiency transmit proportionally larger packets, when they get exclusive access to the WiFi spectrum. Thus, the share of a link decreases with load.

We will say that a hotspot network is \emph{feasible} if it satisfies the constraint that all nodes in the network get at least the rate they got in the baseline network. We want a feasible hotspot network that maximizes the sum of rates of nodes. To exemplify, consider the \emph{baseline} network in Figure~\ref{fig:example1}. It consists of the nodes $A$, $B$, and $C$, each of which uses its link to the tower. As there are three nodes in Figure~\ref{fig:example1}, node $B$ with $30$ dB SINR\footnote{Given our time sharing model, nodes connected to the tower do not interfere with each other. We use SINR to capture the fact that there may be interference due to cells adjacent to the one of our interest and this impacts a node's spectral efficiency.} gets a rate of $(1/3) \log_2(1 + 1000) = 3.32$ bits/sec/Hz. The sum rate is $(1/3) \log_2(1 + 1000) + (2/3) \log_2(1 + 10)$ = $5.63$ bits/sec/Hz.



Let's organize the baseline network into the hotspot network in Figure~\ref{fig:example1a}. Node $B$ with $30$ dB link SINR is chosen as a hotspot and nodes $A$ and $C$ connect as its WiFi clients. All access to the Internet is now via the hotspot $B$'s link, the only used link to the tower. Therefore, the maximum achievable sum rate is the Shannon rate $\log_2(1 + 1000) = 9.96$ bits/sec/Hz of node $B$'s link to the tower. This rate is larger than the sum of the baseline rates of the nodes. To simplify assume the WiFi links are capable of supporting any rate (detailed modeling in Section~\ref{chapter:OptimizationProblem}). Thus each node can be allocated a rate on the hotspot's link to the tower that is larger than its baseline rate. For example, Node $B$ is allocated a rate of $4$ bits/sec/Hz and the remaining $5.96$ bits/sec/Hz is split between $A$ and $C$. The hotspot network in Figure~\ref{fig:example1a} is feasible.


Note that a hotspot network with either node $A$ or $C$, both of whom have a SINR of $10$ dB, as a hotspot instead of node $B$ is infeasible. The maximum achievable sum rate would be $\log_2(1 + 10) = 3.45$ bits/sec/Hz, which is smaller than the sum rate of the baseline network. However, the hotspot network in Figure~\ref{fig:example2a} with nodes $B$ and $C$ as hotspots, and node $A$ as a client of $B$, is feasible, if WiFi links can support any rate. In this network, the link rate of $B$'s link to the tower is $\mathbf{(1/2)} \log_2(1 + 1000) = 4.98$ bits/sec/Hz, which is larger than the sum of the baseline rates of $A$ and $B$. Also, the link rate of $C$'s link to the tower is $(1/2) \log_2(1 + 10)$ bits/sec/Hz, which is greater than its baseline rate. Finally, note that while this hotspot network is feasible, its sum rate is smaller than that of the feasible hotspot network in Figure~\ref{fig:example1a}.

Our specific contributions are:
\begin{itemize}
	\item We formulate the problem of finding an optimal hotspot network configuration. The formulation is a mixed integer non-linear program (MINLP) and takes as input cellular and WiFi link SINR(s). It is described in Section~\ref{chapter:OptimizationProblem}.
	\item We discuss the attributes of the problem in Section~\ref{chapter:attributes} using computationally tractable networks. We demonstrate the interplay between WiFi connectivity between nodes and their cellular SINR(s) in a baseline network and how it impacts the optimal hotspot network. For a couple of network types we find provably optimal hotspot networks (Lemmas~\ref{thm:oneCluster} and~\ref{thm:equalSizeClusters}).
	\item We propose a novel heuristic approach to solve the MINLP, which is detailed in Section~\ref{sec:singleCell}. Our evaluation (Section~\ref{sec:resultsSingleCellTower}) shows that significant rate gains ($20\%$ - $200\%$ in networks we simulated) are available on tethering. We observe that gains are a result of nodes with poorer cellular SINR trading their time share to the tower in the baseline network with nodes that have better SINR, for access to a better SINR link to the tower in the hotspot network.
\end{itemize}

We envisage that the network operator will use the methods proposed herein to configure a network of nodes into a hotspot network. Aspects like impact of mobility, energy consumption, and implementation, are outside the scope of this work. The rest of the paper begins with a description of related works in Section~\ref{sec:related}. We conclude the paper in Section~\ref{sec:conclusions}.
\section{Related Work}
\label{sec:related}
Our work in~\cite{vandanaicc} provides an initial investigation of the problem. Authors in~\cite{6214707} focus on the placement of an a priori fixed number of mobile backbone nodes (MBN) and the assignment of regular nodes, which are mobile, to the MBN(s). 
However, the regular nodes in their network cannot directly connect to the Internet. 
In~\cite{karimi2014optimal} the authors propose an approach that allows both local (to an access point) and non-local users to avail the backhaul available at access points in high access point density wireless LAN(s). The access points are fixed and users are connected to these access points optimally\footnote{Also, see the start-up Velvet \url{http://thisisvest.com/company/velvet/}.}. Instead in our work, the access points (WiFi hotspots) are selected from the given nodes and the other nodes access the Internet via them.

In~\cite{luo2003ucan} the authors design discovery and routing protocols for a network in which peers with high rate links (called proxies) to the cell tower can share their link over a IEEE $802.11$ based ad hoc network with nodes that have low rate links to the tower and are one or more WiFi hops away. The work does not capture the general problem of network reorganization and how it must be done such that it guarantees benefits to all nodes in the network. One of our contributions is to formulate the optimization problem. While we consider only a single hop between a node and its hotspot, we ensure rates of nodes are at least as good as in baseline. In~\cite{Hu_quality_aware_offload}, the authors provide a practical framework of offloading data requests of a user with access to a network (Carrier) with low throughput to another user with access to a network with better throughput. They consider aspects of discovering users, scheduling data of self and others, and motivating users to participate in offloading. Recently in~\cite{azad_wiopt2017}, the authors have extended our proposed model in~\cite{vandanaicc} to also optimize over the time share that a hotspot gets to the tower. They propose variants of a message passing approach to optimize sum rate and to optimize such that rates of nodes increase in proportion to their baseline rates. However, they primarily consider the point coordination function of WiFi under which a hotspot can allocate any time shares to clients.

Works~\cite{ye2013user,nguyen2011efficient} are related to offloading of cellular data to other wireless networks having an independent connection to Internet to achieve load balancing. A user may be offloaded to less congested networks even if the user sees a lower signal-to-interference-and-noise-ratio to these networks. There are other works on offloading cellular data traffic using small cells~\cite{qutqut2013mfw,iosifidis2013iterative,7012044} and WiFi~\cite{siris2013enhancing,lee2010mobile,ding2013enabling,mehmeti_queueing_offloading_model}. Here too the aim is to offload users of the cellular network to other networks that have an independent connection to the Internet. In our work, the connectivity to the Internet is only via cellular.

Device to device (D2D) communication~\cite{lin2014spectrum,pekka2009device,yu2011resource,6855727,golrezaei2014base,asadi2013compound} can enable efficient utilization of the available spectrum. In in-band D2D communication~\cite{lin2014spectrum,pekka2009device,yu2011resource,6855727} the same spectrum is used by both the cellular network and the D2D networks. However, in out-of-band D2D communication~\cite{golrezaei2014base,asadi2013compound} the networks do not share the same spectrum. Similar to out-of-band D2D networks, the WiFi network in our work operates in bands different from that of the cellular network. In~\cite{golrezaei2014base}, authors leverage storage on mobile devices and D2D networks to transfer video files. Authors in~\cite{asadi2013compound} propose to form clusters among cellular users who are in vicinity and can use out-of-band D2D communication. The cluster heads (hotspots) are selected opportunistically. Also, they do \emph{not} ensure that no nodes suffer as a result of cluster formation. Our work can apply to out-of-band D2D communications in which one or more devices provide other devices connectivity to the Internet.

There are works~\cite{bithas2013hybrid,6825080,wang2015energy,sharma2009cool} which focus on optimizing energy consumption by reorganizing the ways in which nodes access the Internet. In~\cite{bithas2013hybrid}, the WLAN access points are wirelessly connected to the eNodeB and share this broadband connection with specific users over WLAN frequencies. This reduces transmission power from eNB to users with bad channel conditions. Authors in~\cite{6825080} consider the case of Internet access to users only via cellular network and divide nodes/mobile users into hotspots and clients in such a way that the overall power consumption is optimized. Authors in~\cite{wang2015energy} propose a scheduling strategy to switch a mobile phone between hotspot and client modes, in order to limit energy consumption. Cool-Tether~\cite{sharma2009cool} takes advantage of the cellular radio links of nearby mobile smart phones to provide energy efficient connectivity by configuring a WiFi hotspot on-the-fly. We do not explicitly optimize energy. However, we expect a hotspot network to consume less energy than the baseline.

Lastly, there are works on self organization that propose clustering algorithms~\cite{heinzelman2002application,liao2013load,lin1997adaptive}. A survey can be found in~\cite{aliu2013survey}.

\section{Optimization Problem}
\label{chapter:OptimizationProblem}
\begin{wrapfigure}{L}{0.3\textwidth}
	\begin{center}
		\includegraphics[width=0.3\columnwidth]{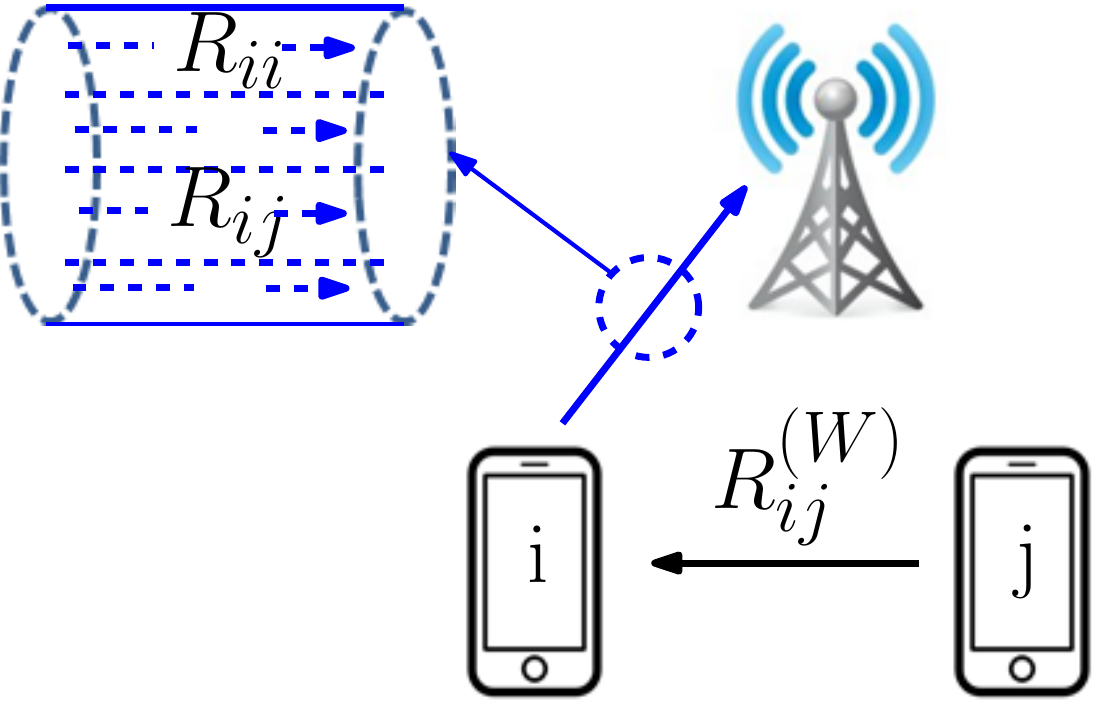}%
		\caption{Node $i$ is a hotspot and node $j$ is its client. We have $a_{ij} = 1$. The rate that node $i$ gets on its link to the tower is $R_{ii}$ and that node $j$ gets on the same link is $R_{ij}$. $R^{(W)}_{ij}$ is the link rate of the WiFi link, defined in~(\ref{eqn:wifiLinkRate}), between client $j$ and hotspot $i$. The inequality $R_{ij} \le R^{(W)}_{ij}$ must be satisfied.}%
		\label{fig:link}%
	\end{center}
		\vspace{-0.2in}
\end{wrapfigure}	
We now formalize our model that we introduced in Section~\ref{chapter:introduction}. Consider a network $\mathcal{N}$ of $N$ nodes, indexed $1,2,\ldots,N$. In the baseline network, node $i$ has a link to the cell tower with a known SINR of SINR$_i$. Each node gets access for a fraction $1/N$ of the total time. The baseline rate $R_i^{(B)}$ of node $i$ is therefore
\begin{align}
	R_i^{(B)} = \frac{1}{N} \log_2(1 + \text{SINR}_i).
	\label{eqn:baselineLinkThr}
\end{align}

Recall that in a hotspot network, a hotspot node $j$ connects to the Internet via its own link to the tower. If the node $j$ is a client, it connects via exactly one other hotspot's link to the tower, the hotspot to which it is assigned. Let $a_{ij}$, $i=1,\ldots,N$, $j=1,\ldots,N$, be variables that indicate assignment of clients to hotspots. We set $a_{ij} = 1$ when $i$ is a hotspot and $j$ is its client. Otherwise, $a_{ij} = 0$. A hotspot $i$ is also its own client, it follows from the above that $a_{ii} = 1$. If $i$ is a client, $a_{ii} = 0$.

Let $R_{ij}$ be the rate allocated for node $j$ on the link to the cell tower of its hotspot $i$. Implicitly, $R_{ii}$ is the rate allocated for hotspot $i$ on its link to the tower. Figure~\ref{fig:link} illustrates how the link of a hotspot node $i$ to the cell tower is shared by itself and node $j$ that is a client of $i$ in a hotspot network. The rate $R_j$ of node $j$ can thus be written as 
\begin{align}
R_{j}=\sum_{i=1}^{N}a_{ij} R_{ij}.
\label{eqn:rateShareNodej}
\end{align}

It follows that if $j$ is a hotspot, then $R_j = R_{jj}$, and if instead $j$ is a client of $i$, then $R_j = R_{ij}$.


\textbf{Modeling the rate of a WiFi link in a hotspot network:} WiFi uses a carrier sense based multiple access with collision avoidance (CSMA/CA)~\cite{bianchi}. The rate at which packets can be serviced by a WiFi link between any two nodes will depend on the received signal strength of the link, the number of co-channel (interfering) hotspots, and the number of nodes connected to the hotspots. Thus the rate at which a WiFi link can service packets depends on the chosen hotspot network configuration and how the hotspots and clients share the available WiFi spectrum. 

In this work, we simplify by assuming that clients of different hotspots don't contend with each other for the medium, either because they are spatially apart or because they are on non-overlapping channels\footnote{Channel allocation is outside the scope of this work. In the $2$ and $5$ GHz band there are, respectively, upto $3$ and $24$ non-overlapping channels.}. Clients of different hotspots may also treat each other's transmission as interference power, the impact of which is captured by the SINR of their links to their respective hotspots. For clients connected to a hotspot, we approximate the impact of the CSMA/CA based channel access mechanism via an efficiency factor $0< \eta \le 1$ (a parameter we vary in simulation). Specifically, $\eta$ captures the fraction of time for which successful packet transmissions take place, under the assumption that the WiFi links are saturated\footnote{See, for example, Figure $6$ in~\cite{bianchi} that plots saturation throughput (defined as the fraction of an \emph{average slot} during which transmissions are successful).}. The medium sees packet collisions or no transmissions for a fraction $1-\eta$ of time. An ideal channel access mechanism under which the medium always sees successful packet transmissions would have $\eta = 1$. Let $\SINR_{ij}$ be the SINR of the WiFi link between nodes $i$ and $j$. Let $\WIFI{ij}$ be the WiFi link rate between the nodes \emph{in a hotspot network}. We define this rate for when $j$ is a client of hotspot $i$. We have
\begin{align}
\WIFI{ij} = 
\begin{cases}
\eta \frac{a_{ij}}{\sum_{k\ne i} a_{ik}}\log_2(1+\SINR_{ij}) & \ \forall i,j \text{ s.t. } i \ne j, a_{ij}=1,\\
\text{undefined} & \text{otherwise}.
\end{cases}
\label{eqn:wifiLinkRate}
\end{align}
When node $j$ is a client of $i$, its link rate is the product of the Shannon rate $\log_2(1+\SINR_{ij})$ of the link and the fraction of time $j$ sends successfully (gets exclusive access to the medium) to its hotspot $i$. This fraction is $\eta \frac{a_{ij}}{\sum_{k\ne i} a_{ik}}$, where $\sum_{k\ne i} a_{ik}$ is the number of clients connected to hotspot $i$ over WiFi.

We seek to maximize the sum rate $\sum_{j=1}^N R_j$ of the network, while ensuring that all nodes get at least their baseline rates (constraint~(\ref{eqn:opt_constraint1}) below). The optimization problem is
\begin{align}
	&\text{Maximize:}\quad \sum_{j=1}^{N}\sum_{i=1}^{N}a_{ij} R_{ij}, \label{eqn:opt}\\
	&{\text{subject to:}}\quad R_{j}\geq R_{j}^{(B)} \quad \forall j,\label{eqn:opt_constraint1}\\
	&{R_{ij} \le \WIFI{ij}}  \quad \forall i,j \text{ s.t. } i \ne j, a_{ij}=1,\label{eqn:opt_constraintWiFi}\\
	&\sum_{j=1}^{N}R_{ij}a_{ij}=\frac{a_{ii}}{H}\log_{2}(1+\SINR_{i}) \quad \forall i,\label{eqn:opt_constraint2}\\
	&\sum_{i=1}^{N}a_{ij}=1 \quad \forall j,\label{eqn:opt_constraint3}\\
	& a_{ij}\in\{0,1\} \quad\forall i,j,\label{eqn:opt_constraint4}
\end{align}
where the (initially unknown) number of hotspots $H=\sum_{i=1}^N a_{ii}$. The variables of optimization are $R_{ij}$ and $a_{ij}$ for all $(i,j)$. The cellular SINR(s), SINR$_i$, $i=1,\ldots,N$, and the WiFi link SINR(s), $\SINR_{ij}$, for any two nodes $i,j$, are inputs to the optimization problem.

Constraint~(\ref{eqn:opt_constraintWiFi}) states that the rate allocated to a client on its hotspot's link to the tower is not larger than the link rate~(\ref{eqn:wifiLinkRate}) of the WiFi link between the client and the hotspot. This ensures that a client $j$ can saturate its rate share $R_{ij}$ to the tower over its WiFi link to its hotspot $i$. Constraint~(\ref{eqn:opt_constraint2}) enforces that all clients of a hotspot get rates such that their sum equals the rate of the hotspot's link to the cell tower. Constraint~(\ref{eqn:opt_constraint3}) enforces that a client must be connected to exactly one hotspot and that a hotspot is not connected to another hotspot. Constraints~(\ref{eqn:opt_constraint1}),~(\ref{eqn:opt_constraint2}),~(\ref{eqn:opt_constraint3}) enforce that clients are not connected to other clients, i.e., if $a_{jj} = 0$ and $a_{kk} = 0$, then $a_{jk} = 0$.

Some preliminary observations on the utility in~(\ref{eqn:opt}) are apparent.  As $a_{ii} = 1$ if $i$ is a hotspot and $a_{ii} = 0$ otherwise, we are choosing hotspots such that the sum of rates of their links to the tower is maximized, while all nodes get at least their baseline rate. Also, if the $a_{ij}$ are set such that the link rate to the tower of node $i$ is greater than or equal to the sum of the baseline rates of all nodes $j$ for which $a_{ij} = 1$, then for such $j$, $R_{ij}$ can be chosen to satisfy~(\ref{eqn:opt_constraint1}) and~(\ref{eqn:opt_constraint2}).
\begin{figure*}[t]
\begin{center}
\subfloat[]{\includegraphics[height=.15\columnwidth]{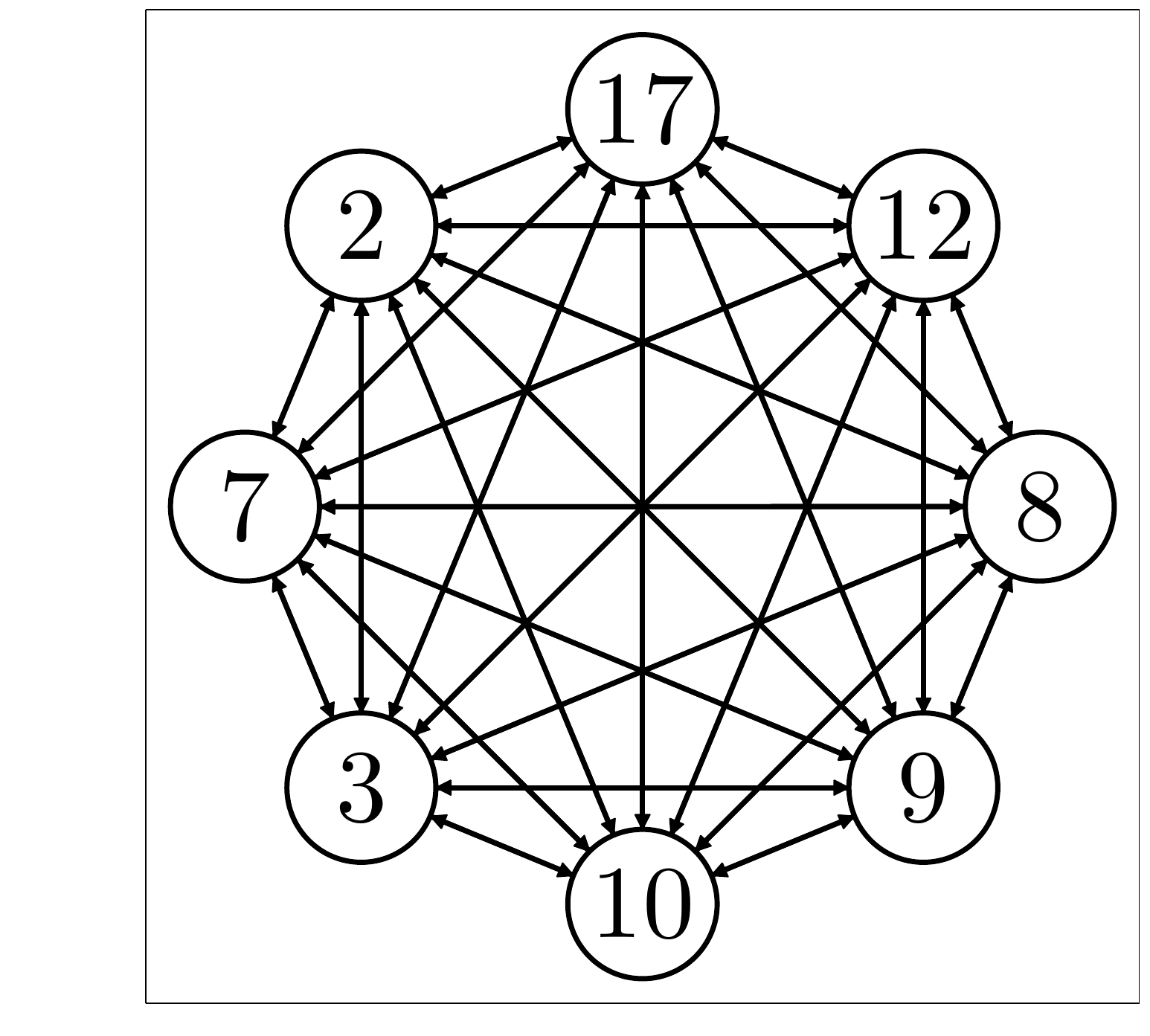}%
\label{fig:network0wifi}%
}\hfill
\renewcommand{\thesubfigure}{a'}
\subfloat[]{\includegraphics[height=.15\columnwidth]{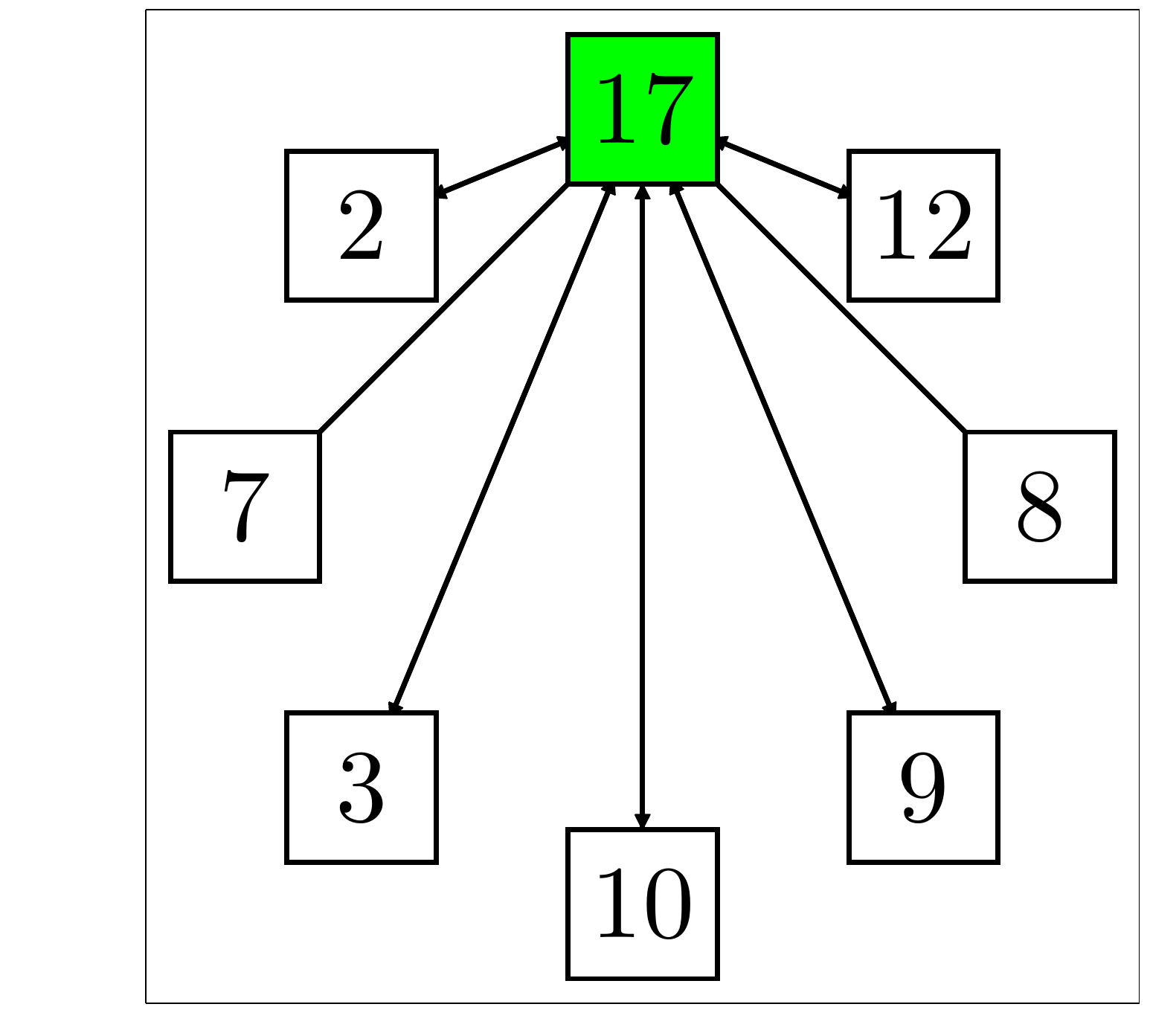}%
\label{fig:network0sol}%
}\hfill
\renewcommand{\thesubfigure}{b}
\subfloat[]{\includegraphics[height=.15\columnwidth]{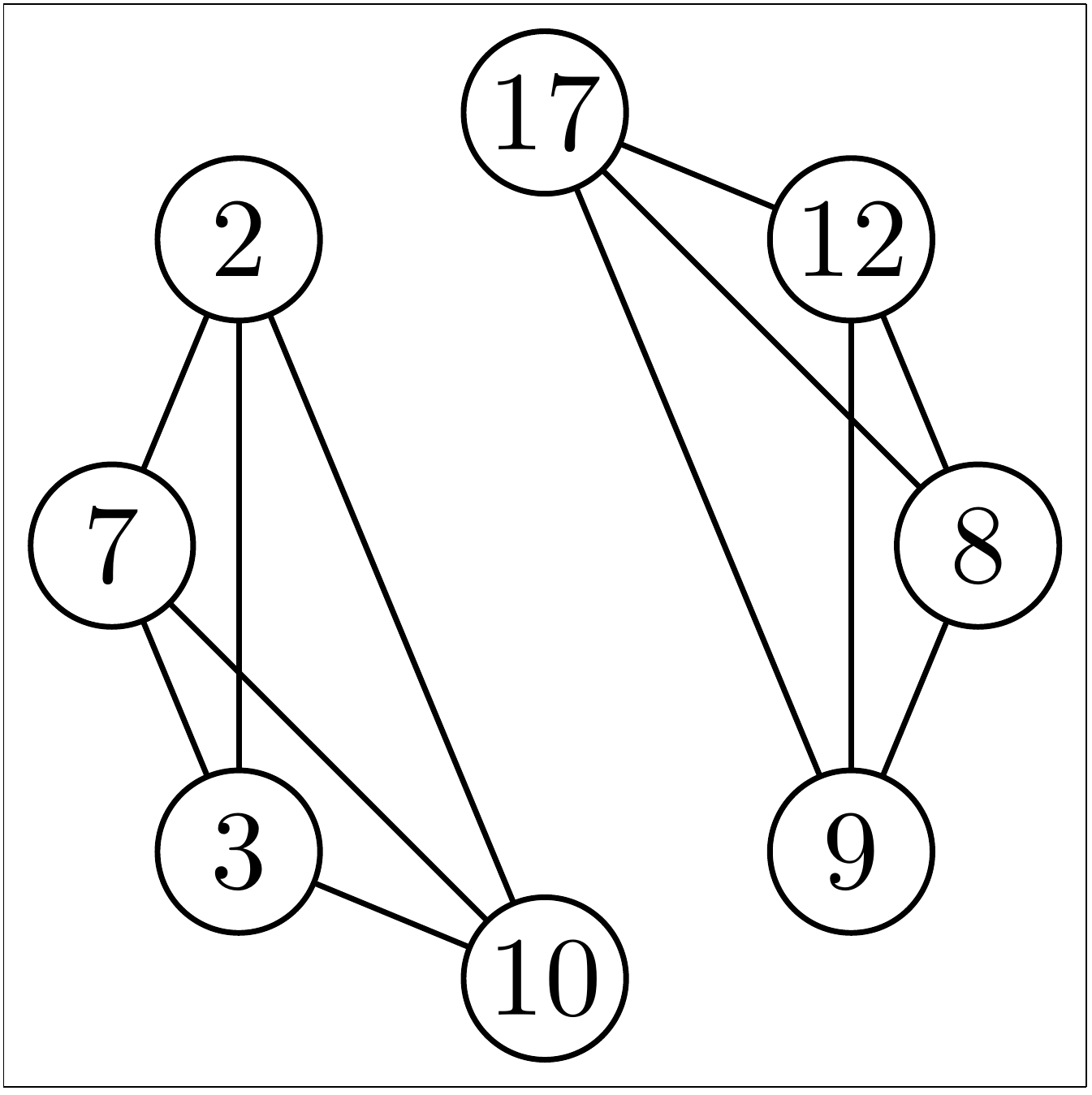}%
\label{fig:network0wifiv1}%
}\hfill
\renewcommand{\thesubfigure}{b'}
\subfloat[]{\includegraphics[height=.15\columnwidth]{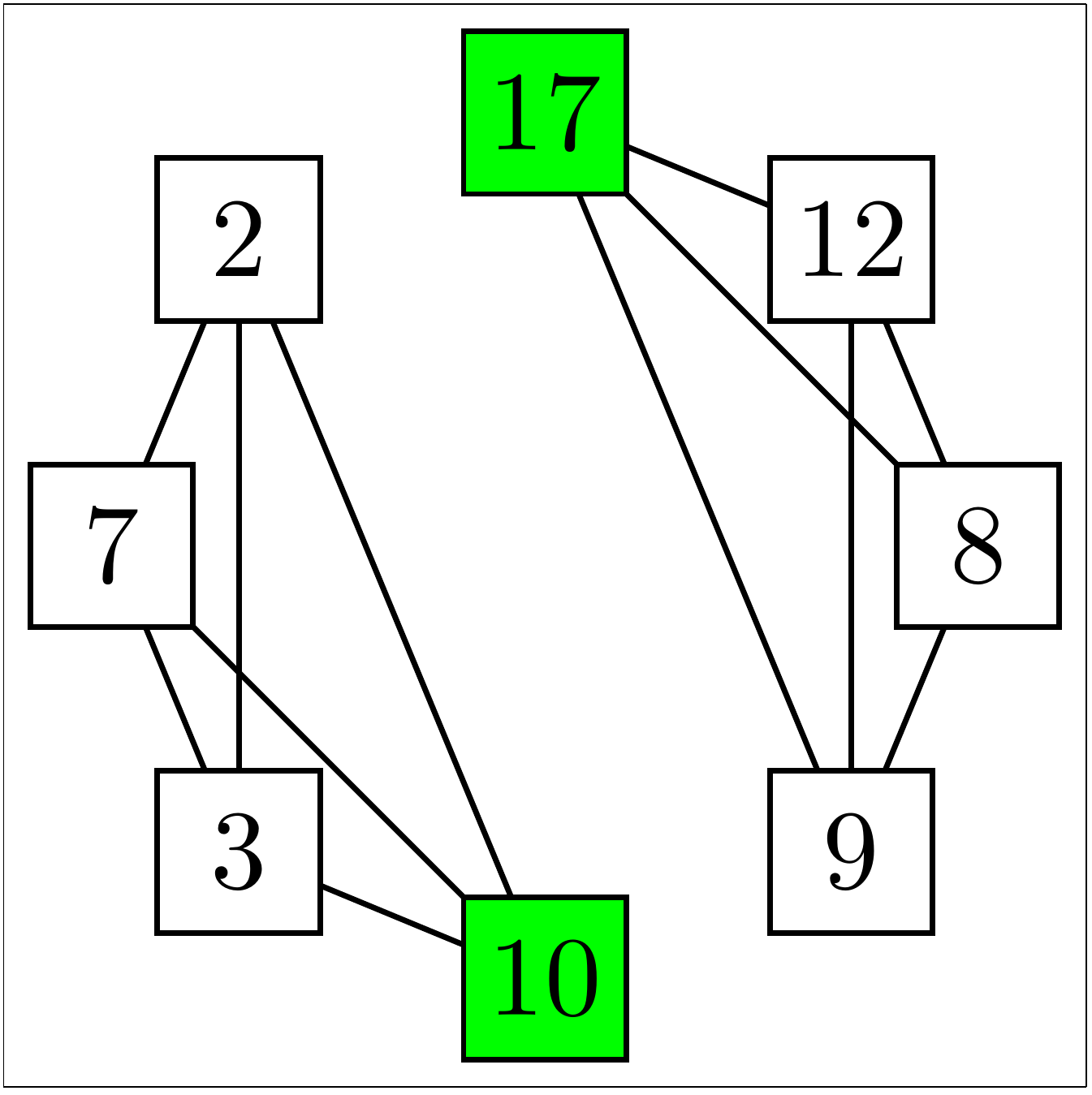}%
\label{fig:network0solv1}%
}\hfill
\renewcommand{\thesubfigure}{c}
\subfloat[]{\includegraphics[height=.15\columnwidth]{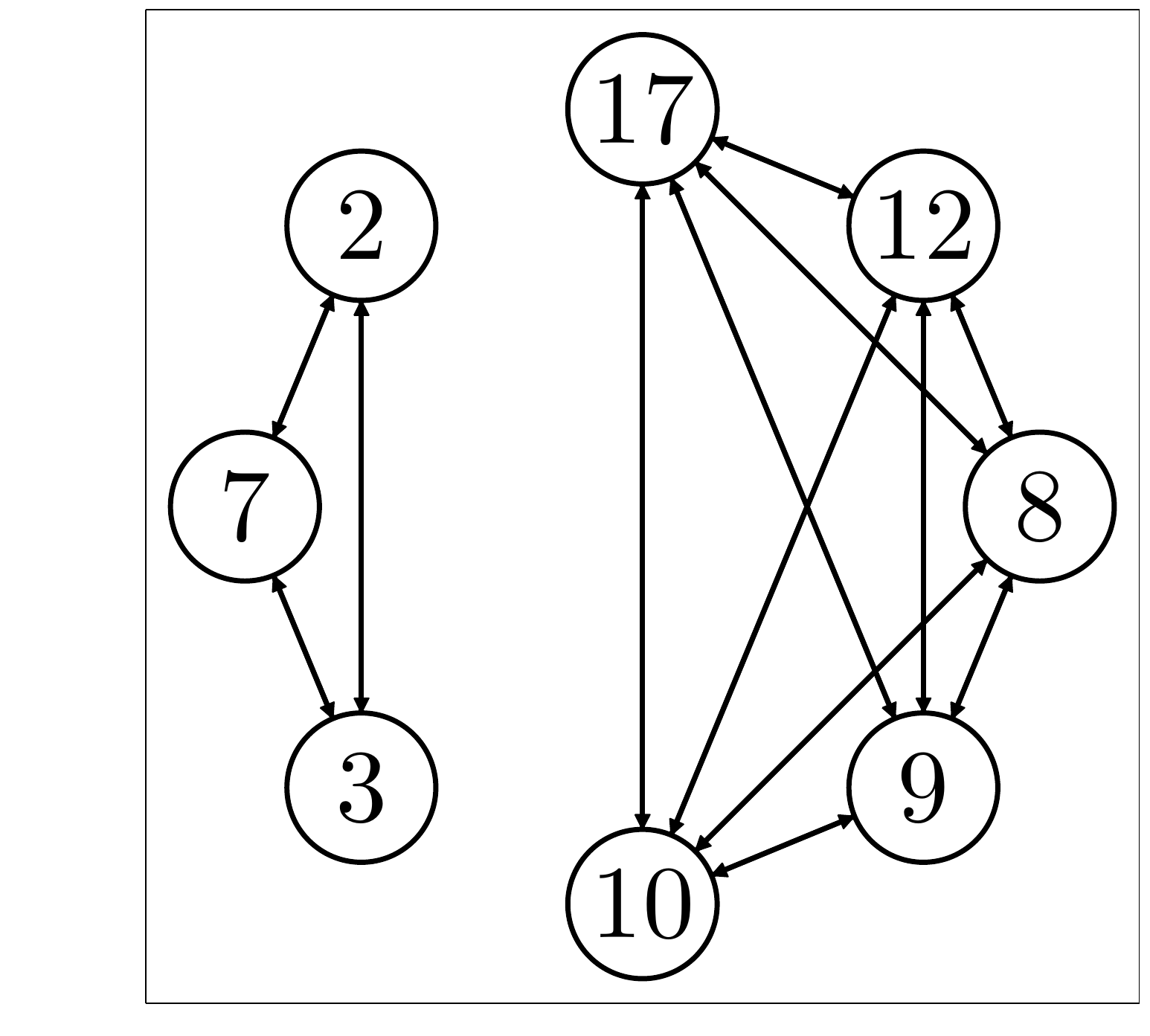}%
\label{fig:network1wifi}%
}\hfill
\renewcommand{\thesubfigure}{c'}
\subfloat[]{\includegraphics[height=.15\columnwidth]{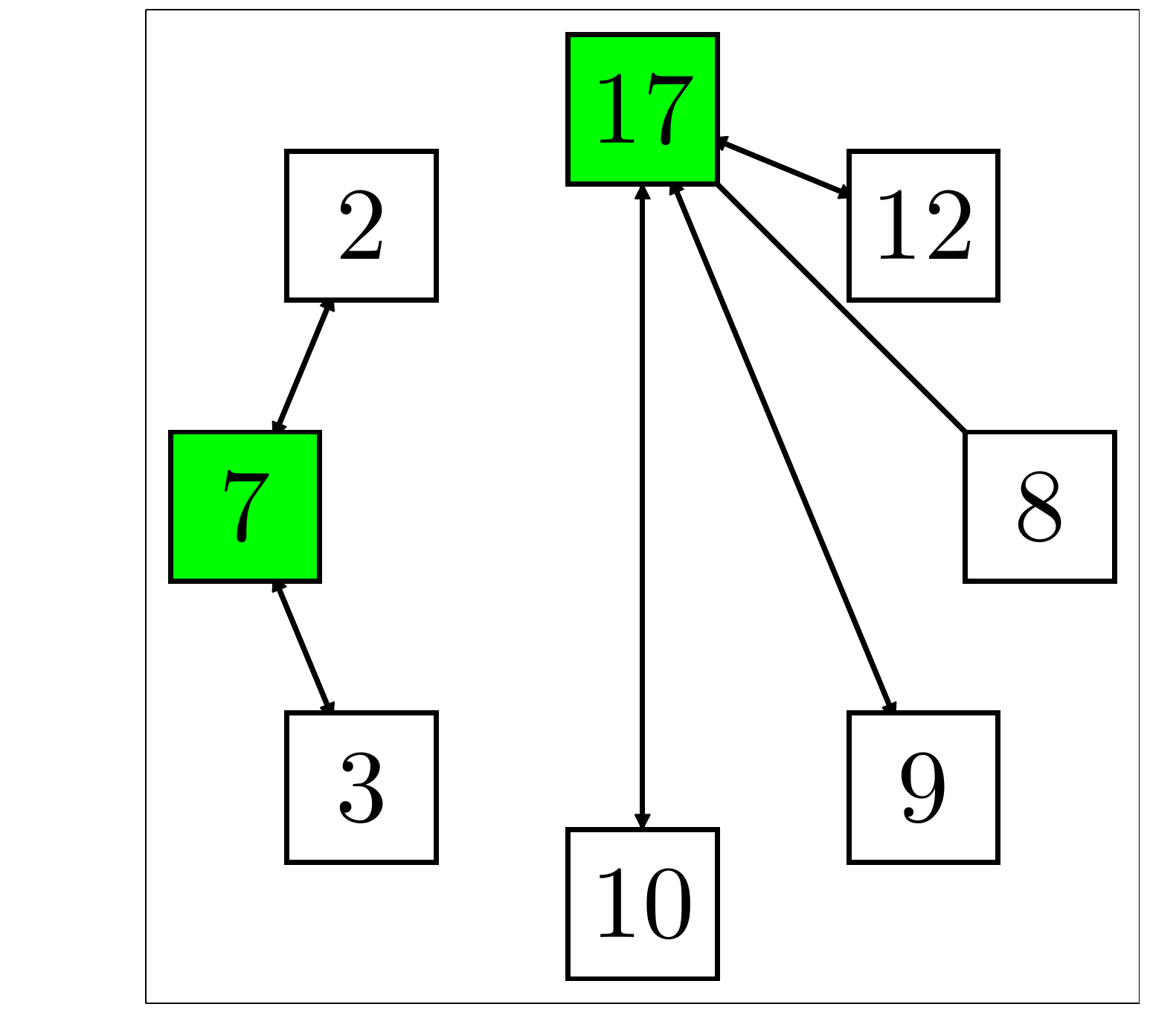}%
\label{fig:network1sol}%
}\hfill
\renewcommand{\thesubfigure}{d}
\subfloat[]{\includegraphics[height=.15\columnwidth]{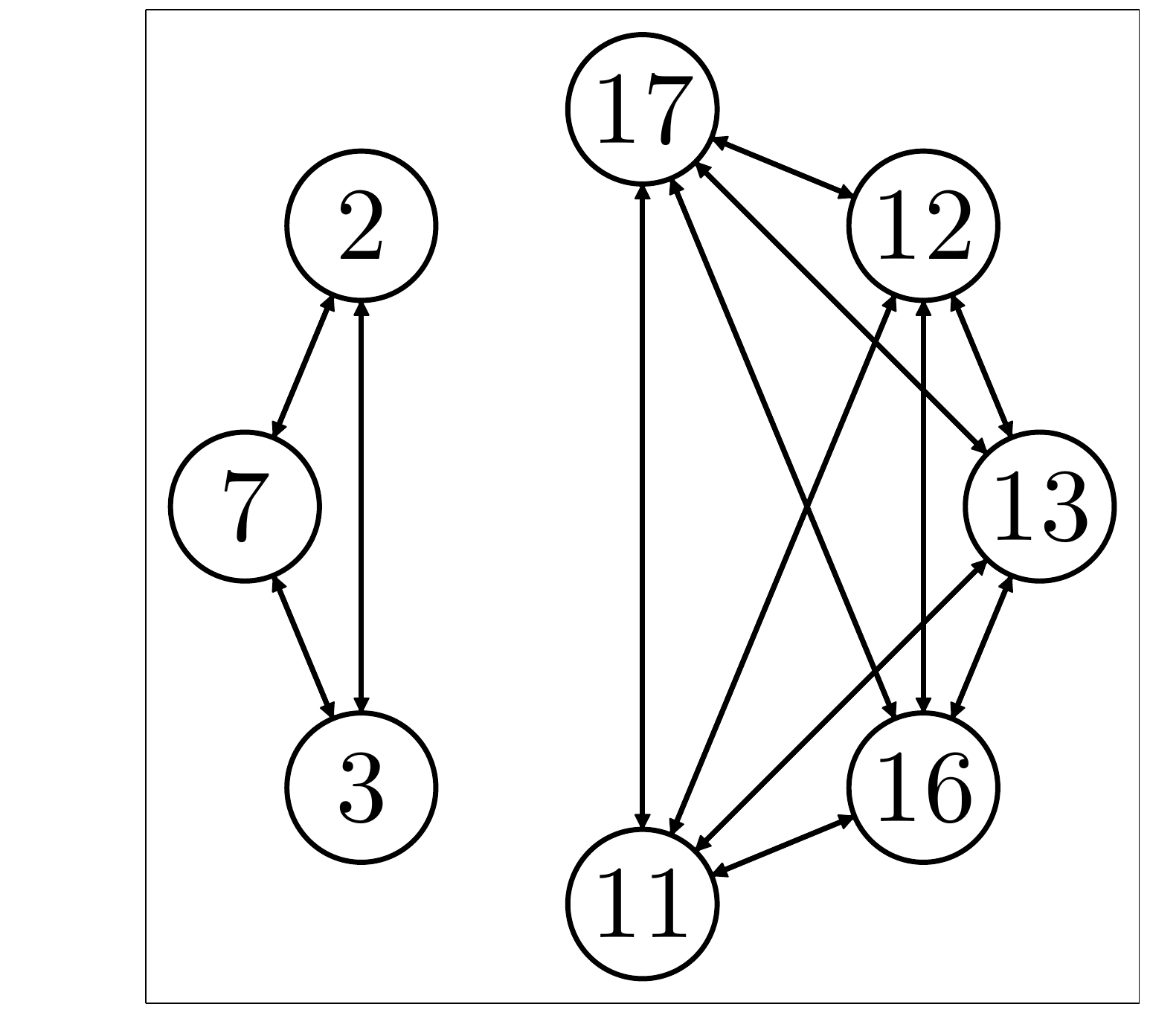}%
\label{fig:network2wifi}%
}\hfill
\renewcommand{\thesubfigure}{d'}
\subfloat[]{\includegraphics[height=.15\columnwidth]{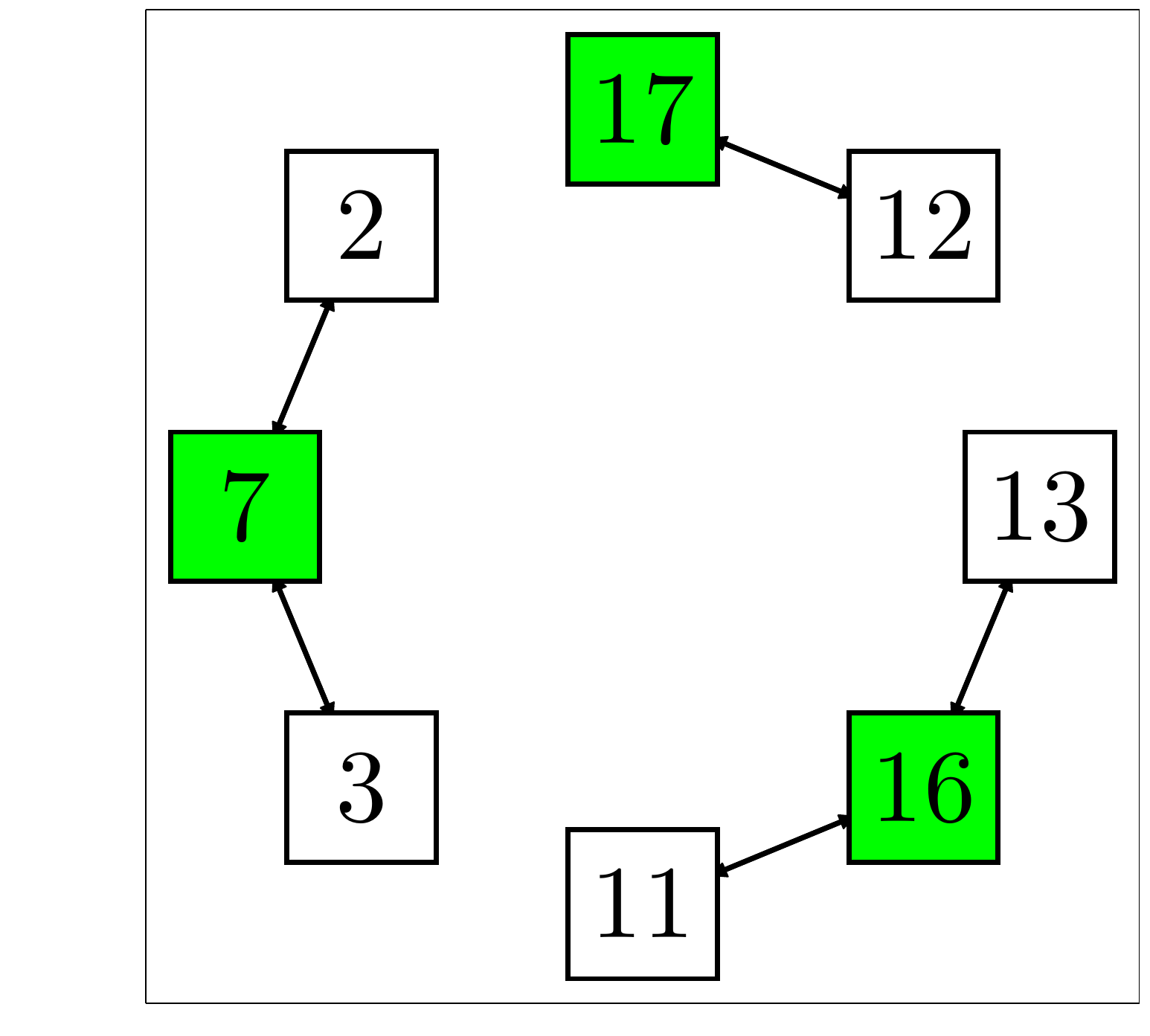}%
\label{fig:network2sol}%
}\hfill
\renewcommand{\thesubfigure}{e}
\subfloat[]{\includegraphics[height=.15\columnwidth]{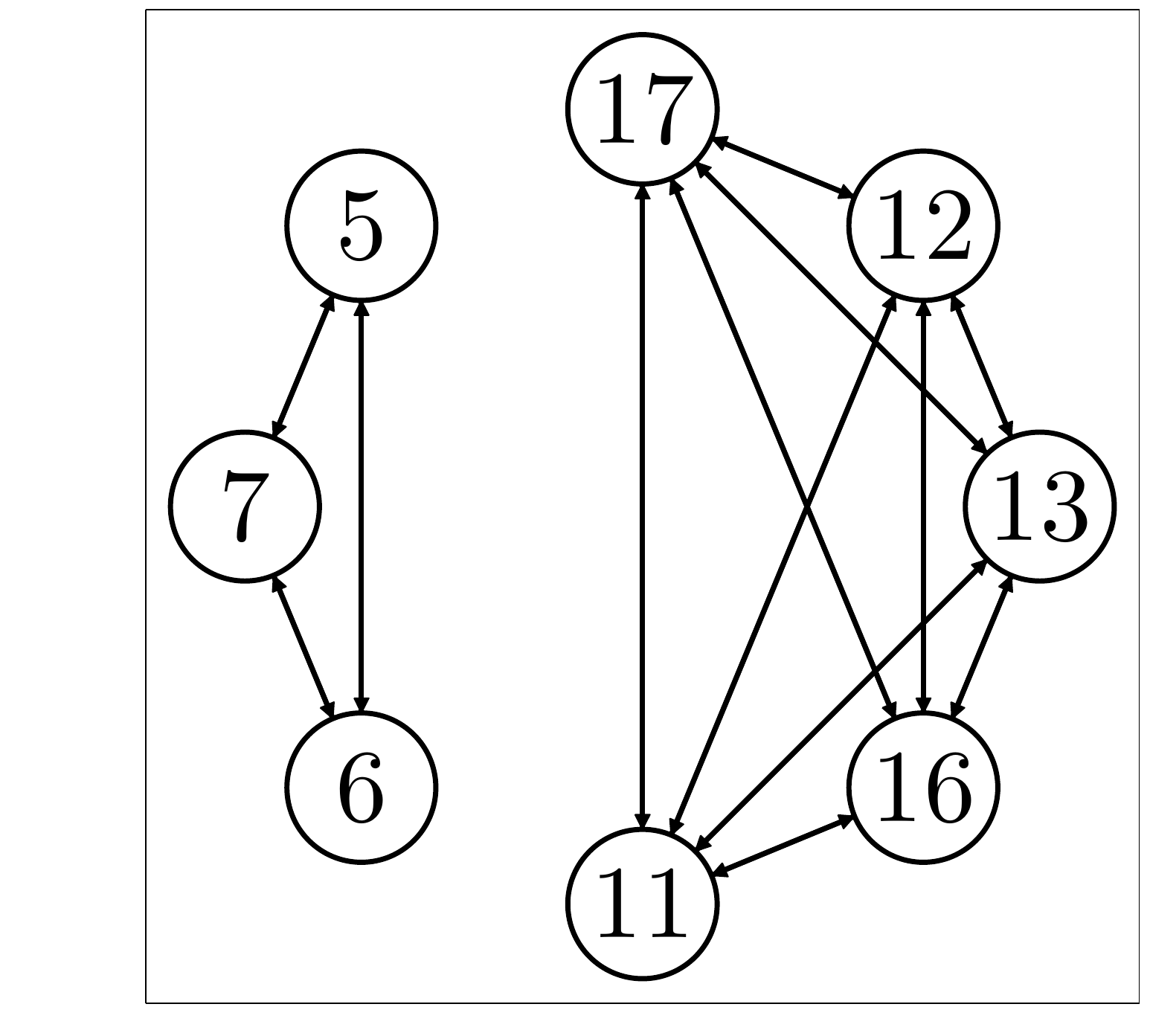}%
\label{fig:network3wifi}%
}\hfill
\renewcommand{\thesubfigure}{e'}
\subfloat[]{\includegraphics[height=.15\columnwidth]{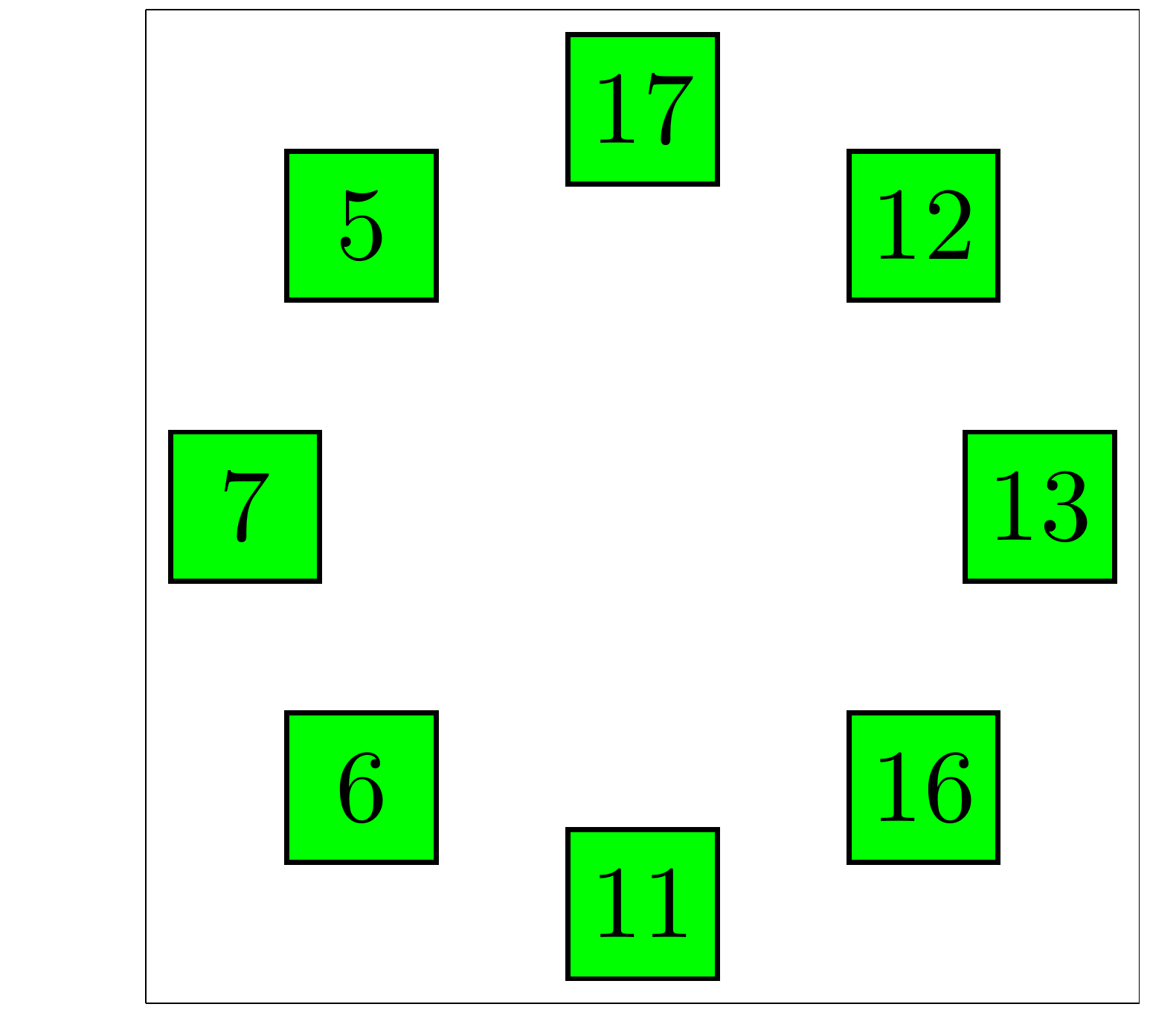}%
\label{fig:network3sol}%
}\hfill
\caption{\small Examples of baseline networks (Figure~\ref{fig:networks}(x)) and the corresponding optimal hotspot networks (Figure~\ref{fig:networks}(x')). For the baseline networks, we show the WiFi network graphs. In them, the nodes are represented by circles. Also, node labels are cellular SINR(s). Optimal hotspot networks are shown by graphs with square nodes. The nodes shaded green are the hotspots. Edges connect a hotspot to its clients.}
\label{fig:networks}
\end{center}
	\vspace{-0.3in}
\end{figure*}

\section{Problem Attributes}
\label{chapter:attributes}
The problem~(\ref{eqn:opt})-(\ref{eqn:opt_constraint4}) is a mixed integer non-linear program. The combinatorial nature of choosing the assignment variables $\{a_{ij}\}$ makes the problem difficult to solve. We illustrate the characteristics of the problem and its solution, using small tractable networks initially, for some insight into the nature of solutions, which were obtained by an exhaustive search. This will motivate our heuristic approach for the general problem in Section~\ref{sec:singleCell}. We will also identify a couple of network types (Lemmas~\ref{thm:oneCluster} and~\ref{thm:equalSizeClusters}) for which the global optimal solution of the problem can be arrived at in a straight forward manner.

Only in this section, we will approximate the WiFi network between nodes by a graph $G=(V_W,E_W)$, where $V_W$ is the set of vertices and $E_W$ is the set of undirected edges. The vertices are the nodes in the network ($V_W = \mathcal{N}$). Denote an edge between $i$ and $j$ by $(i,j)$. If there is an edge between nodes $i$ and $j$, that is $(i,j) \in E_W$, then the WiFi link between $i$ and $j$ is able to service packets at a rate that saturates any rate share $R_{ij}$ (resp. $R_{ji}$) that results from a feasible hotspot network configuration of the nodes in $\mathcal{N}$. Constraint~(\ref{eqn:opt_constraintWiFi}) is always satisfied for such nodes $i$ and $j$, irrespective of $R_{ji}$ and $R_{ij}$. We say that such a link is \emph{not} a \emph{rate bottleneck}. 

If $(i,j) \notin E_W$, then the WiFi link between $i$ and $j$ is only able to service a rate of zero and is a \emph{rate bottleneck}. In Figure~\ref{fig:link} node $j$ is a client of node $i$. Thus, the edge $(i,j) \in E_W$. We will assume that the WiFi network graph $G$ is known for the given network $\mathcal{N}$ of nodes.

The graphs with \emph{circular vertices}, see Figures~\ref{fig:network0wifi}-\ref{fig:network3wifi}, are example WiFi graphs $G$. Also, the labels of nodes are their cellular SINR(s). These graphs thus define different baseline networks. To exemplify, node $10$ in Figure~\ref{fig:network1wifi} has a cellular SINR of $10$ dB. Its WiFi link to node $17$ is \emph{not} a rate bottleneck and that to node $2$ is a rate bottleneck. These graphs are inputs to our optimization problem. The graphs with \emph{square vertices}, see Figures~\ref{fig:network0sol}-\ref{fig:network3sol}, show the optimal hotspot networks corresponding to the baseline networks in Figures~\ref{fig:network0wifi}-\ref{fig:network3wifi}. Vertices shaded green, for example, node $17$ in Figure~\ref{fig:network0sol}, are the hotspots. Edges connect clients to their hotspots.

\emph{Network of nodes for which the WiFi graph $G$ is a clique:} 
Recall that a clique is a fully connected sub-graph. Figure~\ref{fig:network0wifi} shows a WiFi network graph $G$ that is a clique. Thus, none of the WiFi links in the network are rate bottlenecks. Baseline rates $R_i^{(B)}$ of the nodes can be calculated using equation~(\ref{eqn:baselineLinkThr}). The baseline sum rate is $3.0976$ bps/Hz.

Consider the unconstrained optimization of the sum rate~(\ref{eqn:opt}). Specifically,~(\ref{eqn:opt_constraint1})-(\ref{eqn:opt_constraint4}) may not be satisfied. The sum rate is maximized when the node (indexed $17$) with the largest cellular SINR accesses the cell tower all the time. This rate is $\log_2(1 + 10^{(17/10)}) = 5.68$ bits/sec/Hz.

Now observe that the choice of $17$ as the only hotspot is in fact a feasible solution of our constrained problem~(\ref{eqn:opt})-(\ref{eqn:opt_constraint4}). The resulting hotspot network is globally optimal. Specifically, since no WiFi links are bottlenecks, all nodes can be assigned as clients to $17$. We have $a_{17,j} = 1$ for all $j$ and $a_{ij} = 0$ for all $j$ when $i\ne 17$. Also, $R_{ij} = 0$ for all $i\ne 17$ and the number of hotspots $H=1$. For $i=17$, the $R_{ij}$ for all $j$ can be chosen such that $\sum_{j \in \mathcal{N}} R_{ij} = \log_2(1 + 10^{(17/10)})$ (constraint~(\ref{eqn:opt_constraint2})) and $R_{ij} \ge R_{j}^{(B)}$ (constraint~(\ref{eqn:opt_constraint1}), note $R_{ij} = R_j$). An example assignment of $R_{ij}$ is as follows:  reserve a net rate of $R_j^{(B)}= 3.0976$ bits/sec/Hz (sum baseline rate) for all nodes, with individual rates set to baseline rates. Split what remains, $\log_2(1 + 10^{(17/10)}) - 3.0976 = 2.59$ bits/sec/Hz, equally among all nodes. The resulting rates are, in increasing order of node index, $0.5$,$0.5$,$0.6$,$0.7$,$0.7$,$0.8$,$0.8$, and $1.0$ bits/sec/Hz. All nodes see rates larger than their baseline rates. We will use this method of assignment throughout this paper. 

The percentage gain in sum rate obtained by the hotspot network over baseline is $83\%$. The gains can be explained by the fact that the nodes in the hotspot network get access to the Internet via a link with the largest SINR while getting the same cumulative time to the tower as in baseline. Our observations regarding the network in Figure~\ref{fig:network0wifi} are generalized by Lemma~\ref{thm:oneCluster}.

\begin{lemma}\label{thm:oneCluster} Consider a network of $N$ nodes such that the WiFi network $G$ amongst them is a clique. Then there exists an optimal hotspot network that has exactly one hotspot; this hotspot is the node with the largest SINR to the cell tower amongst all nodes.
\end{lemma}

\begin{Proof}\,
The proof can be obtained by reasoning as we did for the network in Figure~\ref{fig:network0wifi}. Specifically, first consider the unconstrained problem and then use the fact that none of the WiFi links are rate bottlenecks.
\end{Proof}

\emph{When the WiFi graph $G$ consists of more than one pairwise disjoint cliques of the same size:}  
The WiFi graph in Figure~\ref{fig:network0wifiv1} has two cliques $C_1 = \{8,9,12,17\}$ and $C_2 = \{2,3,7,10\}$ of equal number of nodes. The cliques are disjoint, that is nodes in different cliques do not have an edge between them. The unconstrained maximization of the sum rate~(\ref{eqn:opt}) is still achieved by selecting $17$ to access the cell tower all the time. However, a hotspot network with only $17$ as a hotspot is infeasible, as nodes in $C_2$ will get rate shares of zero. This is because the WiFi links between $C_2$ and $17$ are bottlenecks.

Now consider the maximization of the sum rate~(\ref{eqn:opt}) under only the constraint that each clique must have at least one link to the tower. The sum rate is maximized by selecting node $17$ from $C_1$ and node $10$ from $C_2$ and equals $0.5 \log_2(1 + 10^{(17/10)}) + 0.5 \log_2(1 + 10^{(10/10)}) = 4.57$ bits/sec/Hz. 
It turns out that a hotspot network with $17$ and $10$ as hotspots and nodes in $C_1$ as clients of $17$ and those in $C_2$ as clients of $10$ is feasible, and thus also optimal. 
These observations are generalized by the following lemma.
\begin{lemma}\label{thm:equalSizeClusters}
Let the WiFi network graph $G$ be such that it can be grouped into $K>1$ pairwise disjoint cliques containing an equal number ($N/K$) of nodes. A hotspot network in which the node with the largest SINR in each clique is chosen as a hotspot is optimal.
\end{lemma}

\begin{Proof}\,
Let nodes be indexed $1,\ldots,N$ and the $K$ cliques $1,\ldots,K$. Let $i^*_k$ be the node with the largest SINR in clique $k$. Consider maximizing the sum rate under only the constraint that each clique must have at least one link to the cell tower. Using arguments made above for the network in Figure~\ref{fig:network0wifiv1}, we can see that the sum rate is maximized by allowing only the links $i^*_k$, $1\le k\le K$, to connect to the cell tower. This maximum sum rate is $\sum_{k=1}^K (1/K) \log_2(1 + \text{SINR}_{i^*_k})$.

Consider a hotspot network in which these $K$ nodes are the hotspots. The node $i^*_k$ that serves as a hotspot for nodes in clique $k$ will get a total rate of $(1/K) \log_2(1 + \text{SINR}_{i^*_k})$ to the tower. For the hotspot network to be feasible, all nodes in clique $k$ must be able to get a rate greater than or equal to their baseline rate via hotspot $i^*_k$. Let every node in a given clique get equal time share over the selected hotspot. Note that a clique gets a total time share of $1/K$. Therefore, each node in clique $k$ must get a time share of $(1/K)/(N/K) = 1/N$, where $N/K$ is the number of nodes in clique $k$. This is exactly the same as the share the nodes got in the baseline configuration. Every node in the clique $k$ now gets a rate of $(1/N) \log_2(1 + SINR_{i^*_k})$. This rate satisfies all the constraints. Therefore, the hotspot network is feasible.
\end{Proof}

The proof highlights the reason why for equal sized cliques, a single hotspot per clique with the node with highest SINR as the hotspot is optimal; the nodes get the same time share as in baseline but access the tower via a link with the highest SINR.

\emph{WiFi network graph $G$ consists of different sized cliques:} 
We look at examples of the more general case (Figures~\ref{fig:network1wifi}-\ref{fig:network3wifi}) in which the cliques are of different sizes. 

Consider the network in Figure~\ref{fig:network1wifi}. The WiFi graph consists of two disjoint cliques $C_1 = \{2,3,7\}$ and $C_2 = \{8,9,10,12,17\}$. Note that the constraints~(\ref{eqn:opt_constraint1})-(\ref{eqn:opt_constraint4}) can be satisfied only if we have two or more hotspots (at least one per clique). Figure~\ref{fig:network1sol} shows the optimal hotspot network configuration, arrived at via an exhaustive search, which has nodes with SINR(s) $7$ dB and $17$ dB as hotspots. Both hotspot selections have maximum SINR(s) within their cliques. In the optimal network (Figure~\ref{fig:network1sol}), hotspot $17$ and its clients ($C_2$) get half the cellular time resource. In the baseline configuration they got a larger share of $5/8$ of the time. Hotspot $7$ and its clients ($C_1$) get half the time instead of $3/8$ of time they got under baseline. The time share of $C_2$ reduces. However, the SINR of $17$ is large enough to satisfy baseline requirements of all nodes in $C_2$. The time share of $C_1$ increases and the nodes now have access to a larger SINR link of node $7$. Thus the sum rate of the network increases. In fact, no other feasible hotspot selections can do as well. This is because any other feasible selection of hotspots will involve either more hotspots (smaller time shares to the tower) and/or hotspots with smaller SINR(s).

The above result in which the optimal hotspot network is obtained by choosing the highest SINR node in each clique is \emph{not} true in general.
Consider  Figure~\ref{fig:network2wifi} that slightly modifies the network in Figure~\ref{fig:network1wifi}. While the WiFi network graph remains unchanged, the SINR(s) of the nodes in the larger clique $C_2$ have a \emph{smaller spread}. 
The hotspot configuration in Figure~\ref{fig:network1sol} can no longer satisfy all the constraints. Specifically, it is easy to verify that the rate of node $17$ to the cell tower, when it gets access to the tower half the time as in Figure~\ref{fig:network1sol}, is smaller than the sum of the baseline rates of all the nodes in its clique. That is $\sum_{j\in C_2} R_{j}^{(B)} > 0.5 \log_2(1 + \text{SINR}_{17})$. Nothing has changed with respect to the smaller clique $C_1$ and so node $7$ can act as a hotspot for it as long as $7$ gets half the time.

Since, given half the time, $17$ is not a feasible hotspot for the larger clique, none of the other nodes in the clique, given that their SINR(s) are smaller than $17$, can be the only hotspot for the larger clique. It turns out that in the optimal network the three nodes $17$, $16$, and $7$ are hotspots. In this network, the clique containing $17$ gets $2/3$ of the time as it has two hotspots $17$ and $16$, instead of the $1/2$ it got when $17$ was chosen to be the only hotspot for the clique. So while on an average the nodes in $C_2$ have access to a smaller SINR link ($16$ dB and $17$ dB instead of all accessing a link with $17$ dB SINR), they have access to a larger time share. Also, $7$ as a hotspot can support the baseline rates of the nodes in the smaller clique, even when it gets only $1/3$ of the time. This hotspot network is shown in Figure~\ref{fig:network2sol}.


Finally, consider the network in Figure~\ref{fig:network3wifi}. The WiFi graph retains the larger clique of Figure~\ref{fig:network2wifi}. However, its smaller clique has nodes with larger cellular SINR(s) and a smaller spread in the SINR(s). It turns out that the baseline configuration (see Figure~\ref{fig:network3sol}) is in fact sum rate optimal. That aside, it is instructive to consider the feasibility of choosing two hotspots $6$ and $7$ for the smaller clique and three hotspots $13$, $16$, and $17$ for the larger clique. Each hotspot gets $1/5$ of the time. It turns out that the sum rate of hotspots of both the cliques is greater than the sum of baseline rates of the nodes in them. However, no feasible assignment of nodes in the smaller clique to the hotspots $6$ and $7$ exists\footnote{For a feasible assignment to exist, we must be able to assign $5$ to one of $6$ and $7$. At $1/5$ of the time, node $7$ has a rate of $0.518$ bits/sec/Hz to the tower and node $6$ has a rate of $0.463$ bits/sec/Hz. The baseline rates of $5,6,7$ are $0.257, 0.29, 0.324$ bits/sec/Hz respectively. Node $5$ cannot be assigned to $7$ (and hence $6$) as the link of $7$ to the tower has a rate of $0.518$ bps, which is less than the sum $0.58$ bps of baseline rates of $7$ and $5$.}.


In summary, nodes with good WiFi connectivity amongst each other can not be optimized independently of others, as configuring a group of nodes into hotspots and clients impacts the possible configurations of other nodes. For a node to be a hotspot for other nodes, not only its cellular SINR, but also the SINR of the other nodes, and the total share of time the nodes get in the hotspot network configuration are important.
\section{Greedy Heuristic Approach}
\label{sec:singleCell}
\begin{algorithm}
	\small
	Input: ${\mathcal{N}}$, $\SINR_{ij}$, $\SINR_i$, for all $i,j\in \mathcal{N}$;
	Output: Hotspot Network Configuration\\
	Compute $C$ (equation~(\ref{eqn:cij})); $R^{(B)}_i$ (equation~(\ref{eqn:baselineLinkThr})), for all $i\in \mathcal{N}$;\\	
	$W = \textbf{Prospective-Client-Matrix}(C,R^{(B)}_1,\ldots,R^{(B)}_N)$;
	
	$\text{Compute} \  i_1,i_2,....,i_N \ s.t. \ \text{SINR}_{i_1} \geq \text{SINR}_{i_2}....\geq \text{SINR}_{i_N} $ \\
	\For{$H = 1 \text{ to } N$}{			
		
		$\mathcal{H} = \emptyset$;\ $\mathcal{U}_h = \emptyset,\ 0\le h\le H$;
		$\mathcal{H}' = \emptyset$; $\mathcal{R}' = \mathcal{N}$; $c=1$\\
		\While{${\mathcal{R}'} \ne \emptyset$\ $\&\&$\ $c\le H$}{		
			$\mathcal{H} = \mathcal{H} \cup \mathcal{H}'(1)$;\quad
			$\mathcal{U} = \cup_{h=c}^H \mathcal{U}_h$;\quad
			$\mathcal{N}' = \mathcal{R}' \cup \mathcal{U}$;\\
			$[\mathcal{H}',{\mathcal{R}'},\mathcal{U}_c,\dots,\mathcal{U}_H] = \textbf{Select Hotspots}(\mathcal{N}',W,H,H - c + 1)$; 
			$c= c + 1$;
		}		
		$\mathcal{H} = \mathcal{H} \cup \mathcal{H}'$;
		$R_H = 0$;\\
		\uIf{${\mathcal{R}'} == \emptyset$}{
			$R_H = \frac{1}{H} \sum_{i\in \mathcal{H}} \log_2(1 + \SINR_i)$;
			save $\mathcal{H},\mathcal{U}_1,\dots,\mathcal{U}_H$ as the $H$-hotspot network.
		}
		\tcp{Exit condition.}
     \If{$H<|{\mathcal{N}}|$ \text{AND} $\frac{1}{H+1} \sum_{i\in i_1,...,i_{H+1}} \log_2(1 + \SINR_i) < R_{H}$}{
			break;}								
	}
  $H^* = \arg\max_{h\in \{1,\ldots,H\}} R_h$;\\
  
  \textbf{Compute-Fair-Loading}($H^*$-hotspot network);\tcp{Algorithm~\ref{alg:fairLoading}}
  \tcp{\textbf{Compute $R_{ij}$ to satisfy constraint~(\ref{eqn:opt_constraint2}) for all $i,j$ such that $a_{ij} = 1$}}
  Return $H^*$-hotspot network;\\
	\caption{\textbf{Configure-Network}}
	\label{alg:confNetwork}	
\end{algorithm}
Algorithm~\ref{alg:confNetwork} (\textbf{Configure-Network}) encapsulates our approach. A step wise summary is next.
\begin{enumerate}
	\item Algorithm~\ref{alg:createWiFiAdj} (\textbf{Prospective-Client-Matrix}): For every node, fix the set of all other nodes in the network that \emph{may} become the node's clients. The set is summarized as a \emph{prospective client matrix} $W$. It is obtained using WiFi link SINR(s) and the baseline rates of nodes. Later, we pick clients of a node $i$ \emph{only} from this set. This ensures that any client $j$ of node $i$ satisfies equation~(\ref{eqn:opt_constraintWiFi}) in any hotspot network with node $i$ as a hotspot.
	\item Look for a feasible hotspot network with exactly $H$ hotspots, for $H\in \{1,\ldots,N\}$. Start with $H=1$. Do the below listed for each $H$, till $H\le N$ or the \emph{exit condition} (in Algorithm~\ref{alg:confNetwork}) is satisfied. From the obtained networks, pick the $H$-hotspot network with the largest sum rate. For each $H$ do the following (Algorithm~\ref{alg:selectHotspots} (\textbf{Select-Hotspots})):
	\begin{enumerate}
	\item For each node, calculate the set of nodes that can be its clients using a simple cellular \emph{SINR-based heuristic}, under the assumption that the node is one of $H$ hotspots (has a fraction $1/H$ of the time to the tower). In addition to~(\ref{eqn:opt_constraintWiFi}), this set satisfies~(\ref{eqn:opt_constraint1}).
	\item Greedily pick $H$ nodes as hotspots in decreasing order of the number of clients they can support, such that they together provide connectivity to all nodes in the network. If a selection of hotspots is found, it satisfies~(\ref{eqn:opt_constraint1}),~(\ref{eqn:opt_constraintWiFi}),~(\ref{eqn:opt_constraint3}), and~(\ref{eqn:opt_constraint4}).
	\end{enumerate}
	\item Algorithm~\ref{alg:fairLoading} (\textbf{Compute-Fair-Loading}): Reassign clients to the hotspots in the chosen hotspot network such that the hotspots are \emph{fairly} loaded. This leads to larger per node rate gains, for the same gains, over baseline, in the sum rate of the network.
	\item Allocate rates $R_j$ to all nodes in the hotspot network such that Equation~(\ref{eqn:opt_constraint2}) is satisfied. Such an allocation is guaranteed to exist for a hotspot network obtained as above. An example allocation was shown in Section~\ref{chapter:attributes}.
\end{enumerate}
\begin{algorithm}
		\small
		Input: $C$, $R^{(B)}_i$, for $i \in \mathcal{N}$\\
		Output: $W = (w_{ij})$\\
		\For{$i \in \mathcal{N}$}{
		Initialize $w_{ij} = 0$, for $j  \in \mathcal{N}$;\\		
		Calculate $S_i^{(n)} = \{j\ne i: c_{ij}/n \ge R_j^{(B)}\}$, $n=1,\ldots,N$;\\	
		Set $n^*$ to largest $n$ for which $|S_i^{(n)}| \ge n$;\\ 
		\uIf {$n^*$ \text{exists}}{
		Compute $i_1,\ldots,i_{|S_i^{(n^*)}|}$ such that $R^{(B)}_{i_1} < \ldots < R^{(B)}_{i_{|S_i^{(n^*)}|}}$;\\
		Set $w_{ij} = 1$, for $j=i_1,\ldots,i_{(n^*)}$;
		}
		}
Return $W$;\\	
\caption{\textbf{Prospective-Client-Matrix}}
\label{alg:createWiFiAdj}	
\end{algorithm}
\vspace{-0.3in}

\begin{algorithm}
\small
\begin{multicols}{2}
	Input: ${\mathcal{N}'}$, $W$, $H$, $H'$\\
	Output: $\mathcal{H}',{\mathcal{R}'},\mathcal{U}_1,\dots,\mathcal{U}_{H'}$ 
	\\$W' = W$;
	$\mathcal{H}'=\emptyset$;\\
	\For{$i \in {\mathcal{N}'}$}{		
		$\mathcal{O} = \{o\in \mathcal{N}': w'_{io} = 1, o\ne i\}$;\\
		$R = R^{(B)}_i + \sum_{j\in \mathcal{O}} R^{(B)}_j$;\quad
		$A = \frac{1}{H} \log_2(1 + \SINR_i)$;\\
		\tcp{The SINR-based heuristic.}
		\While{$A < R$}{		
			$k^* = \arg\max_{k\in \mathcal{O}} \SINR_k$; \\
			$w'_{ik^*} = 0$;\tcp{Remove $k^*$ from prospective clients of $i$.}
			$\mathcal{O} = \{o: w'_{io} = 1, o\in \mathcal{N}'\}$; 
			$R = R^{(B)}_i + \sum_{j\in \mathcal{O}} R^{(B)}_j$;\\
		}			  
	}
	\tcp{Pick $H'$ nodes as hotspots.}
	\For{$h = 1 \text{ to } H'$}{
		$i^* = \arg\max_{n \in {\mathcal{N}'}} |\{k: w'_{kn} = 1\}|$; \tcp{If more than one node maximizes the function (number of supported clients), pick $i^*$ to be the node with the largest SINR.}
		$\mathcal{H}' = \mathcal{H}' \cup i^*$;\\
		$\mathcal{U}_h = \{u:w'_{i^{*}u} = 1\}$; 
		${\mathcal{N}'} = {\mathcal{N}'} - \mathcal{U}_h$;
	}
	$\mathcal{R}' = \mathcal{N}'$;\tcp{The set of unassigned nodes.}
	\caption{\textbf{Select-Hotspots}}
	\label{alg:selectHotspots}
	\end{multicols}
\end{algorithm}
\vspace{-0.3in}

\begin{algorithm}
	\small
	Input and Output: Set of hotspots $\mathcal{H}$ and their clients.\\
	Calculate $\mathcal{F}$ for current hotspot network.\\
	$\delta\mathcal{F} = \mathcal{F}$;\tcp{Change in fairness index}
	\While{$\delta\mathcal{F} \ge 0$}{	
		$i^{*} = \arg \min_{i\in \mathcal{H}} (L_i)$;\\
		$E_{i^{*}} = \frac{\log_2 (1 + \text{SINR}_{i^{*}})}{H} - \sum_{j = 1}^{N} {R_{j}^{(B)} a_{i^{*}j}}$ ;\tcp{Remaining capacity of minimally loaded hotspot}
		$A_{i^{*}} = \{j: w_{i^{*}l} = 1, j\notin \mathcal{H}, a_{ij} \ne 1, R_{j}^{(B)} < E_{i^{*}}\}$;\\			
		$Q = \{i: a_{ij} = 1, j\in A_{i^{*}}\}$;\tcp{Set of hotspots for each node $j\in A_{i^{*}}$}
			
		$q^{*} = \arg \max_{q\in Q} (L_q) $; \tcp{Select hotspot with largest load}
		$u^{*} = \arg\max_{j\in A_{i^{*}}} (a_{q^{*}j}*R^{(B)}_j)$;\tcp{Select client of $q^*$ with largest baseline rate}
		Calculate change in fairness $\delta\mathcal{F}$ in case $u^{*}$ is reassigned to $i^{*}$ as a client;
		
		\uIf{$\delta\mathcal{F} > 0$}
		{
			Set $a_{i^{*}u^{*}} = 1; a_{q^{*}u^{*}} = 0$;\tcp{Modify hotspot network}
		}

	}
	Return updated hotspot network.
	\caption{Compute-Fair-Loading}
	\label{alg:fairLoading}
\end{algorithm}
\vspace{-0.3in}

\subsection{Algorithmic Details} 
\label{sec:algoDetails}
\begin{figure}
	\begin{center}
		\subfloat[\small Cellular SINR(s) in dB (node labels) and the corresponding baseline rates (edge labels) in bits/sec/Hz.]{
		\includegraphics[height=.25\columnwidth]{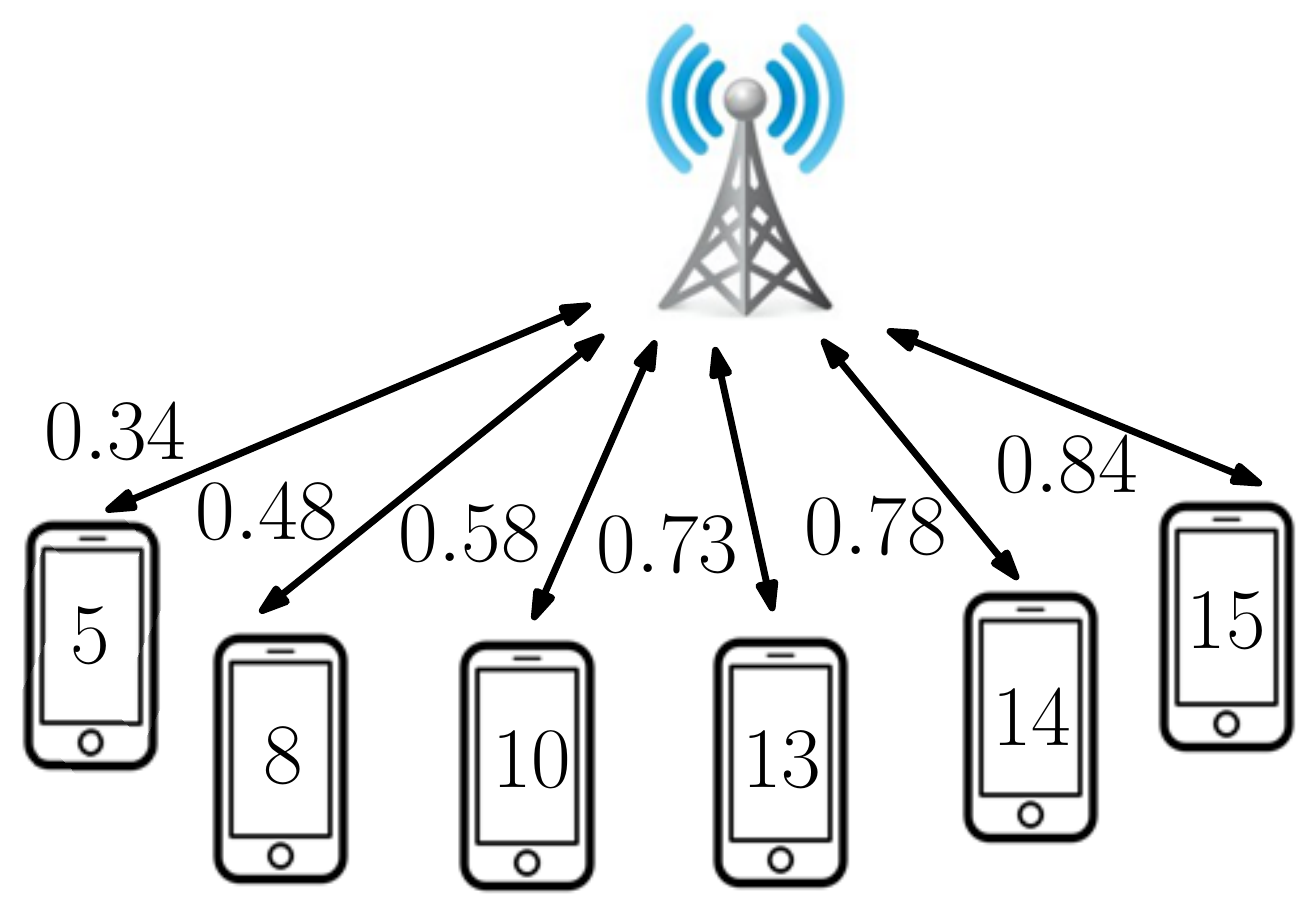}%
			\label{fig:exampleBaselineNtwk}%
		}\hspace{.2in}
		\subfloat[\small Pairwise WiFi SINR(s) $\text{SINR}_{ij}$ dB.]{			
			\small{
				\begin{tabular}[b]{|l|l|l|l|l|l|l|}
					\hline
					$\mathbf{i}$\ / $\mathbf{j}$ & $\mathbf{5}$ & $\mathbf{8}$ & $\mathbf{10}$  & $\mathbf{13}$  & $\mathbf{14}$  & $\mathbf{15}$   \\ \hline
					$\mathbf{5}$ & \textemdash & $-10$  & $-8$  & $29$  & $20$  & $22$ \\ \hline
					$\mathbf{8}$ & $-10$  & \textemdash & $20$  & $25$  & $-8$ & $-5$  \\ \hline
					$\mathbf{10}$& $-8$  & $20$  & \textemdash & $21$  & $-15$  & $-6$  \\ \hline
					$\mathbf{13}$& $29$  & $25$  & $21$  & \textemdash & $22$  & $26$  \\ \hline
					$\mathbf{14}$& $20$  & $-8$ & $-15$  & $22$  & \textemdash & $21$  \\ \hline
					$\mathbf{15}$& $22$ & $-5$  & $-6$  & $26$  & $21$  & \textemdash \\ \hline
				\end{tabular}	
				}							
			\label{tab:exampleWifiSNR}	
		} 
		\end{center}
		\caption{An example network of six nodes. Network $\mathcal{N}=\{5,8,10,13,14,15\}$.}
			\vspace{-0.3in}
\end{figure}

Algorithm~\ref{alg:confNetwork} takes as input the network $\mathcal{N}$, cellular link SINR(s) $\SINR_i$ for all nodes $i$ in the network, and WiFi link SINR(s) $\SINR_{ij}$ between any pair of nodes $i$ and $j$. It first invokes Algorithm~\ref{alg:createWiFiAdj}. We use the example network in Figure~\ref{fig:exampleBaselineNtwk} to exemplify our approach. The pairwise WiFi SINR(s) between nodes in the network are shown in Table~\ref{tab:exampleWifiSNR}.

Algorithm~\ref{alg:createWiFiAdj} details creation of a prospective client matrix $W = (w_{ij}) \in \{0,1\}^{N\times N}$. The matrix restricts, for every node $i$, the nodes $j$ that may be considered as clients of node $i$. The algorithm requires as input the baseline rates and the matrix $C = (c_{ij})$, where 
\begin{align}
c_{ij} = \eta \log_2(1 + \SINR_{ij}).
\label{eqn:cij}
\end{align}

Let the set $S^{(n)}_i$ consist of all nodes $j\ne i$ whose WiFi link rates $\WIFI{ij}$ to node $i$ can support at least their baseline rates when they are among a total of $n$ nodes, chosen from $S^{(n)}_i$, and \emph{assigned as clients to node} $i$. Thus for any node $j\in S^{(n)}_i$, we require $c_{ij}/n \ge R^{(B)}_j$. For such a set to be feasible, it must contain at least $n$ nodes. Let $n^*$  be the largest $n$ that gives us a feasible $S^{(n)}_i$. Since the set $S^{(n^*)}_i$ may contain $|S^{(n^*)}_i| \ge n^*$ nodes, we select a set of $n^*$ nodes that have the smallest baseline rates from it. For every selected node $j$, we set $w_{ij} = 1$. Otherwise, we set $w_{ij} = 0$. If a feasible $S^{(n)}_i$ is not found for any $n$, then $i$ has no prospective clients.

\begin{figure}		
\begin{center}
		\subfloat[\small The table enumerates $c_{ij}/n$ for node $i=15$. For node $15$, $n^* = 3$ and  $S^{(n^*)}_{15} = \{5,13,14\}$.]{
			\centering
			\begin{adjustbox}{width=0.51\columnwidth}
				\small{
				\begin{tabular}[b]{|l|l|l|l|l|l|}

					\hline
					$\mathbf{n}$\ /\ $\mathbf{(i,j)}$ & $\mathbf{(15,5)}$ & $\mathbf{(15,8)}$  & $\mathbf{(15,10)}$  & $\mathbf{(15,13)}$  & $\mathbf{(15,14)}$ \\ \hline
					$\mathbf{1}$     & $\mathit{7.32}$ & $0.40$  & $0.32$  & $\mathit{8.64}$  & $\mathit{6.99}$ \\ \hline
					$\mathbf{2}$    & $\mathit{3.66}$  & $0.20$ & $0.16$ & $\mathit{4.32}$  & $\mathit{3.49}$ \\ \hline
					$\mathbf{3}$    & $\mathit{2.44}$  &  $0.13$ & $0.11$  & $\mathit{2.88}$ &  $\mathit{2.33}$ \\ \hline
					$\mathbf{4}$    & $\mathit{1.83}$  & $0.10$ & $0.08$  & $\mathit{2.16}$  & $\mathit{1.75}$ \\ \hline
					$\mathbf{5}$    & $\mathit{1.46}$  & $0.08$ & $0.06$ & $\mathit{1.73}$  & $\mathit{1.40}$ \\ \hline
				\end{tabular}
				}
			\end{adjustbox}
			\label{tab:wifiCij}
		}
		\hspace{.18in}
\subfloat[\small The WiFi prospective client matrix $W$, shown as a graph with nodes as vertices and a directional edge from node $i$ to $j$ when $w_{ij} = 1$.]{\includegraphics[height=.13\columnwidth]{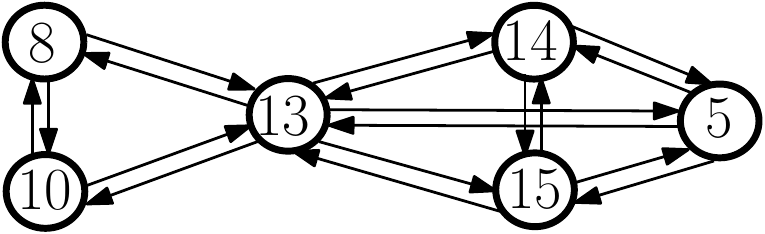}%
\label{fig:exampleProspectiveMatrix}%
}
\caption{\small Exemplification of the workings of Algorithm~\ref{alg:createWiFiAdj} for the network described by Figure~\ref{fig:exampleBaselineNtwk} and Table~\ref{tab:exampleWifiSNR}.}		
\end{center}
	\vspace{-0.3in}
\end{figure}

To exemplify, in Table~\ref{tab:wifiCij} we list $(c_{ij})/n$ for node $i=15$ of the network in Figure~\ref{fig:exampleBaselineNtwk}. For all $n$, only for nodes $j\in \{5,13,14\}$, $c_{15,j}/n$ is greater than the corresponding baseline rates. Thus, for node $i=15$, $n^*=3$. Also, $S^{(n^*)}_{15} = \{5,13,14\}$. Figure~\ref{fig:exampleProspectiveMatrix} shows the prospective WiFi matrix for the network in graph form. The vertices are the nodes. Nodes $i,j$ for which $w_{ij} = 1$ are connected by a directional edge from $i\rightarrow j$, signifying that $j$ is a prospective client of $i$.

\emph{Configure a $H$-hotspot network:} For a given number $H$ of hotspots, let $\mathcal{H}$ be the set of selected hotspots. Let $\mathcal{U}_i$ be the sets of clients of hotspot $i$. Note that $i\in \mathcal{U}_i$. Let $R_{H}$ be the sum rate of the $H$-hotspot network. Let $\mathcal{R}'$ be the set of nodes that were not assigned to any of the $H$ hotspots. For a given $H$, we initialize the sets $\mathcal{H}, \mathcal{U}_1,\ldots,\mathcal{U}_H$ to empty sets and we let $\mathcal{R}'=\mathcal{N}$.

Post Algorithm~\ref{alg:createWiFiAdj}, Algorithm~\ref{alg:confNetwork} attempts to configure a $H$-hotspot network by up to $H$ calls of Algorithm~\ref{alg:selectHotspots}. The need for multiple calls is explained later. The first time it is called for a given $H$, we set $H' = H$ and $\mathcal{N}' = \mathcal{N}$.

Algorithm~\ref{alg:selectHotspots} (Select-Hotspots) takes a given set $\mathcal{N}' \subseteq \mathcal{N}$ of nodes, the target number $H$ of hotspots in the hotspot network, the number $H'$ that remains to be selected, and the matrix $W$. In the algorithm, we decide for every node $i \in \mathcal{N}'$, the nodes in $\mathcal{N}'$ that can become its clients, when it has a time-share of $1/H$ to the cell tower. Let the set $\mathcal{O}$ be initialized to contain all nodes that are prospective clients of $i$ as per the matrix $W$. We use the \emph{SINR-based heuristic} to prune this set so that the sum of the baseline rates of all nodes in the resulting set and that of $i$ are smaller than the rate of $i$'s link to the tower. As per the heuristic, nodes that have a larger cellular SINR are removed first from the set $\mathcal{O}$. This is motivated by the fact that, everything else equal, a node with larger cellular SINR is a more desirable hotspot.

After prospective clients for every node in $\mathcal{N}'$ have been finalized, we look for $H'$ nodes that together provide Internet connectivity to all nodes in $\mathcal{N}'$. The $H'$ nodes are chosen greedily in decreasing order of the number of their prospective clients. In addition to returning the set $\mathcal{H}'$ of hotspots and the corresponding sets of clients, the algorithm also returns the set of nodes $\mathcal{R}'$ that were not covered in the above sets. If the set is nonempty, the selected hotspots do not provide connectivity to all nodes in $\mathcal{N}'$. Hence, the selection is infeasible. This infeasibility may often result from the greedy decision making. As a workaround, Select-Hotspots is called a maximum of $H$ times when looking for a $H$-hotspot network.

In general, the $c\textsuperscript{th}$ time Select-Hotspots is called, it returns $H'=H-c+1$ hotspots in the set $\mathcal{H}'$ and their set of clients $\mathcal{U}_c,\ldots,\mathcal{U}_H$. Prior to the $c\textsuperscript{th}$ call, $c-1$ hotspots and their clients have been selected. All the selected hotspots are stored in $\mathcal{H}$. If $\mathcal{R}' \ne \emptyset$, the hotspot that was selected first amongst the $H'$ hotspots, denoted by $\mathcal{H}'(1)$, and its set $\mathcal{U}_c$ of clients, are added to the existing hotspot network configuration. In the next iteration ($c = c + 1$), the rest of the hotspots and clients (the set $\cup_{h=c}^{H} \mathcal{U}_h$) and the nodes in $\mathcal{R}'$ are sent back to Select-Hotspots. If $\mathcal{R}' \ne \emptyset$ after $H$ calls of Select-Hotspots, a $H$-hotspot network is declared infeasible ($R_H = 0$).

\begin{figure}
\begin{center}
		\subfloat[\small Run of Select-Hotspots for $H=1$.]{\includegraphics[height=.1\columnwidth]{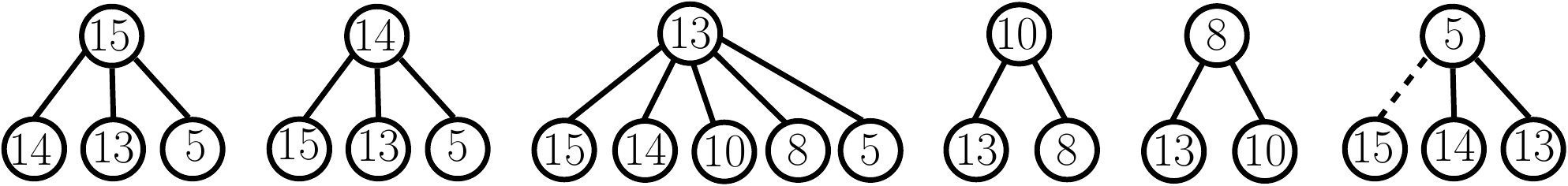}%
			\label{fig:firstStage}%
		}\\
		\subfloat[\small Run of Select-Hotspots for $H=2$.]{\includegraphics[height=.1\columnwidth]{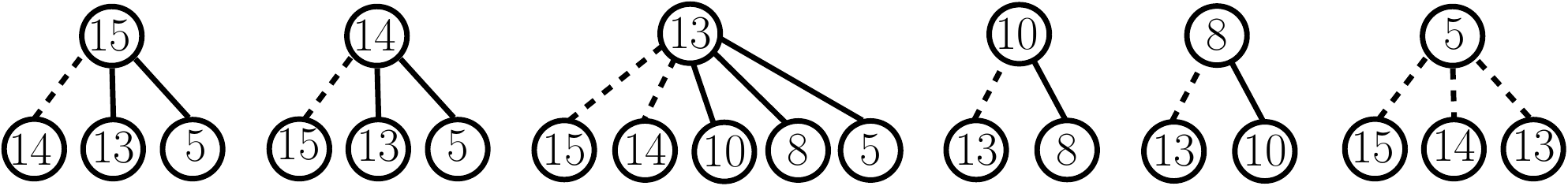}%
			\label{fig:secondStage}%
		}\\
		\subfloat[\small Second run of Select-Hotspots for $H=2$.]{\includegraphics[height=.1\columnwidth]{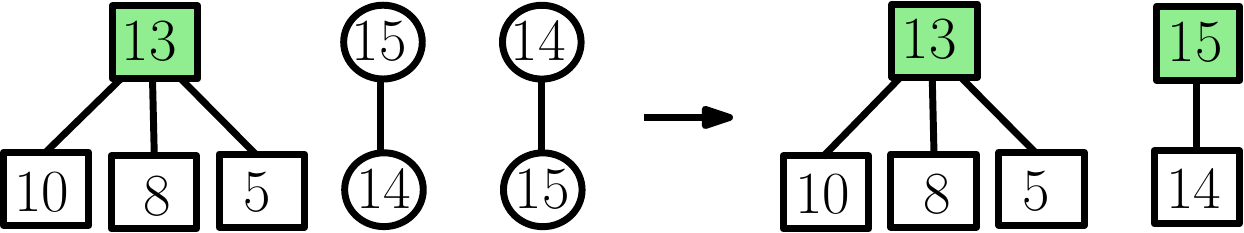}%
			\label{fig:thirdStage}%
		}\hspace{0.2in}
		\subfloat[\small Final hotspot network obtained after Compute-Fair-Loading.]{\includegraphics[height=.1\columnwidth]{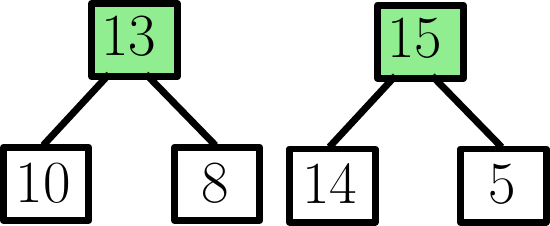}%
			\label{fig:fairnessStage}%
		}
		\caption{\small Result of algorithms Select-Hotspots and Compute-Fair-Loading on the network in Figure~\ref{fig:exampleBaselineNtwk}.}
		\label{fig:algoExampleIter}
	\end{center}
	\vspace{-0.3in}
\end{figure}

To exemplify, for the network in Figure~\ref{fig:exampleBaselineNtwk}, Figure~\ref{fig:firstStage} shows the result of calling Select-Hotspots for $H=1$. Node $5$ has nodes $13$, $14$, and $15$ as its prospective clients given by Algorithm~\ref{alg:createWiFiAdj} and shown in Figure~\ref{fig:exampleProspectiveMatrix}. Clearly, node $5$ can not be the only hotspot in the network. In fact, node $5$ can't support the sum of baseline rates of $13$, $14$, and $15$. Node $5$ as the only hotspot gets a rate of $\log_2(1 + \SINR_5) = 2.057$ bits/second and the sum of baseline rates of $5$, $13$, $14$, and $15$ is $2.69$ bits/second. Using the SINR-based heuristic, node $15$ is removed (dashed line in Figure~\ref{fig:firstStage}) from the list of prospective clients of $5$ for $H=1$. Other nodes are able to support all their prospective clients when $H=1$. In addition, we know from Figure~\ref{fig:exampleProspectiveMatrix} that all nodes are prospective clients of node $13$. Thus, $H=1$ with node $13$ as the hotspot is feasible.

Similarly, the first run of Select-Hotspots for $H=2$ gives nodes and clients they can support as shown in Figure~\ref{fig:secondStage}. Since node $13$ has the largest number of clients, it will be chosen as the first of the two hotspots. The second hotspot must be either node $14$ or $15$. However, neither of them has the other as a prospective client (dashed lines connect them in Figure~\ref{fig:secondStage}). This seeming infeasibility of choosing two hotspots is resolved by calling Select-Hotspots for $H=2$ one more time, over the nodes $15$ and $14$, with the goal of adding one more hotspot ($H=2$, $H'=1$). This gives us Figure~\ref{fig:thirdStage}. To the left of the arrow, we have a hotspot network configuration created in part, and nodes $14$ and $15$ with their feasible clients. While we want to select $H'=1$ hotspot from $14$ and $15$, their available capacity ($A$ in Select-Hotspots) is calculated for the final target of $H=2$ hotspots. To the right of the arrow, we have the final hotspot network configuration. Since $15$ and $14$ have the same connectivity, $15$ was chosen as a hotspot due to its larger SINR.

\emph{Ensuring fairness in loading (Algorithm~\ref{alg:fairLoading}):} After Algorithm~\ref{alg:selectHotspots} has been invoked for $1\le H \le N$, we pick the H-hotspot network that maximizes the sum rate $R_H$ (see Algorithm~\ref{alg:confNetwork}). In Algorithm~\ref{alg:fairLoading}, we rearrange clients amongst the $H$ hotspots of the chosen network such that the hotspots are fairly loaded. Next we define loading $L_i$ of a hotspot $i$ and the fairness index $\mathcal{F}$.
\begin{align}
L_i = \frac{H}{\log_2(1+\SINR_i)} \sum_{j=1}^{N} a_{ij} R^{(B)}_j;\quad
\mathcal{F} = \frac{(\sum_{i\in \mathcal{H}} L_i)^2}{H \sum_{i\in \mathcal{H}} L^2_i}.
\end{align}
Loading $L_i$ is simply the sum baseline rate requirement of nodes using the link to the tower of node $i$ as a fraction of the rate of the link, given that there are $H$ hotspots. The fairness index (Jain's index~\cite{raj_jain}) is calculated over loading(s) of the set of selected hotspots $\mathcal{H}$. Jain's fairness index $0\le \mathcal{F} \le 1$. A larger index implies greater fairness. 

In each iteration of Algorithm~\ref{alg:fairLoading} we pick the hotspot $i^{*}$ with the smallest loading $L_{i^{*}}$. Let $A_{i^{*}}$ be the set of nodes that are clients in the current hotspot network and may be reassigned to $i^{*}$ as its clients. Such clients must be amongst the prospective clients of $i^{*}$, as per $W$. They must not already be clients of $i^{*}$. Their baseline rate requirements must be small enough to be accommodated by $i^{*}$, given the existing clients of $i^{*}$. For each such client, we compute its current hotspot. We pick the most loaded hotspot amongst those obtained. Amongst clients of this hotspot in the set $A_{i^{*}}$, we pick the one that has the largest baseline rate requirement and assign it as a client to $i^{*}$. We repeat this process till it increases fairness.

For our example network in Figure~\ref{fig:exampleBaselineNtwk}, the last run of Select-Hotspots gave the hotspot network in Figure~\ref{fig:thirdStage}. In the network, while $13$ has a smaller rate to the cell tower than $15$, it services a larger sum-baseline requirement. The loading of $13$, $L_{13} = 0.97$, and that of $15$, $L_{15} = 0.64$. Application of Algorithm~\ref{alg:fairLoading} rearranges the clients to give the hotspot network in Figure~\ref{fig:fairnessStage}. Note that the hotspots are still the same and hence the sum-rate of the network stays unchanged. However, now we have a similar loading of $L_{13} = 0.82$ and $L_{15} = 0.78$. This often leads to fairness in per node rate (calculation of rate $R_j$ of node $j$ is explained next) gains over baseline\footnote{For our example, the rate gains of nodes $5,8,10,13,14,15$ before and after fair loading are, respectively, $4.76\%,3.41\%, 2.82\%, 2.23\%, 56.81\%, 53.19\%$ and $53.31\%, 28.43\%, 23.59\%, 18.59\%, 23.30\%, 21.82\%$.}.

The obtained hotspot network ensures that every node gets at least its baseline rate, that is every node $j$ can get a rate $R_j \ge R^{(B)}_j$. We set each node's rate such that Equation~(\ref{eqn:opt_constraint2}) is also satisfied. Specifically, for every hotspot, we start by assigning all its clients their baseline rates. The link rate to the tower that remains unused is further split equally amongst clients. Each client adds to its rate share the entire split or a smaller portion as allowed by its WiFi link rate (see Equations~(\ref{eqn:wifiLinkRate}) and~(\ref{eqn:opt_constraintWiFi})). Repeat, till~(\ref{eqn:opt_constraint2}) is satisfied for the hotspot. Since hotspot $i$ is its own client and is not constrained by a WiFi link, it is always possible to satisfy~(\ref{eqn:opt_constraint2}). 

\emph{Note on the exit condition in Algorithm~\ref{alg:confNetwork}:} Consider the following. If the largest possible sum rate for any selection (feasible or not) of $H+1$ hotspots, which is the sum of the link rates to the tower of nodes with the $H+1$ largest cellular SINR(s), is less than the rate $R_H$ obtained from a feasible selection of $H$ hotspots, then the sum rate maximizing hotspot configuration must contain $\le H$ hotspots. This fact motivates the exit condition in Algorithm~\ref{alg:confNetwork}. The use of the exit condition leads to improvements in the run time of the algorithm in practice\footnote{For networks of $100,\ 200$ and $400$ nodes, the average number of hotspots that Configure-Network considers (maximum value that $H$ takes) before returning a hotspot network configuration are, respectively, $44.87,\ 80.66,\text{ and }152.84$, which are less than half the number of nodes in the network. The use of the exit condition does not change the obtained hotspot network.}.
\section{Evaluation Methodology and Results} \label{sec:resultsSingleCellTower}

We simulated networks of nodes distributed uniformly and randomly over a circular region, with the cell tower at the center. We chose radii of $\mathcal{R}=1,\ 2, \text{ and } 5$ km, number of nodes $N = 100,\ 200, \text{ and }400$, efficiency values of $\eta = 0.5,\ 0.75,\text{ and } 1$, a pathloss exponent of $\alpha = 3$ that is used to model the wireless channel for cellular links and $\alpha_W = 2.5, \text{ and } 3$ for modeling the channel for WiFi links. For each selection of $\mathcal{R}$, $N$, $\eta$, $\alpha$, and $\alpha_{W}$, we simulated $100$ network instances. We show averages over the $100$ instances, unless stated otherwise. The cell tower transmits at a power of $1$ W and is placed at a height of $30$ meters. WiFi transmissions use a power of $100$ mW. Both use a bandwidth of $20$ MHz.


\begin{figure}
	\begin{center}
		\subfloat[\small]{\includegraphics[width=.35\columnwidth]{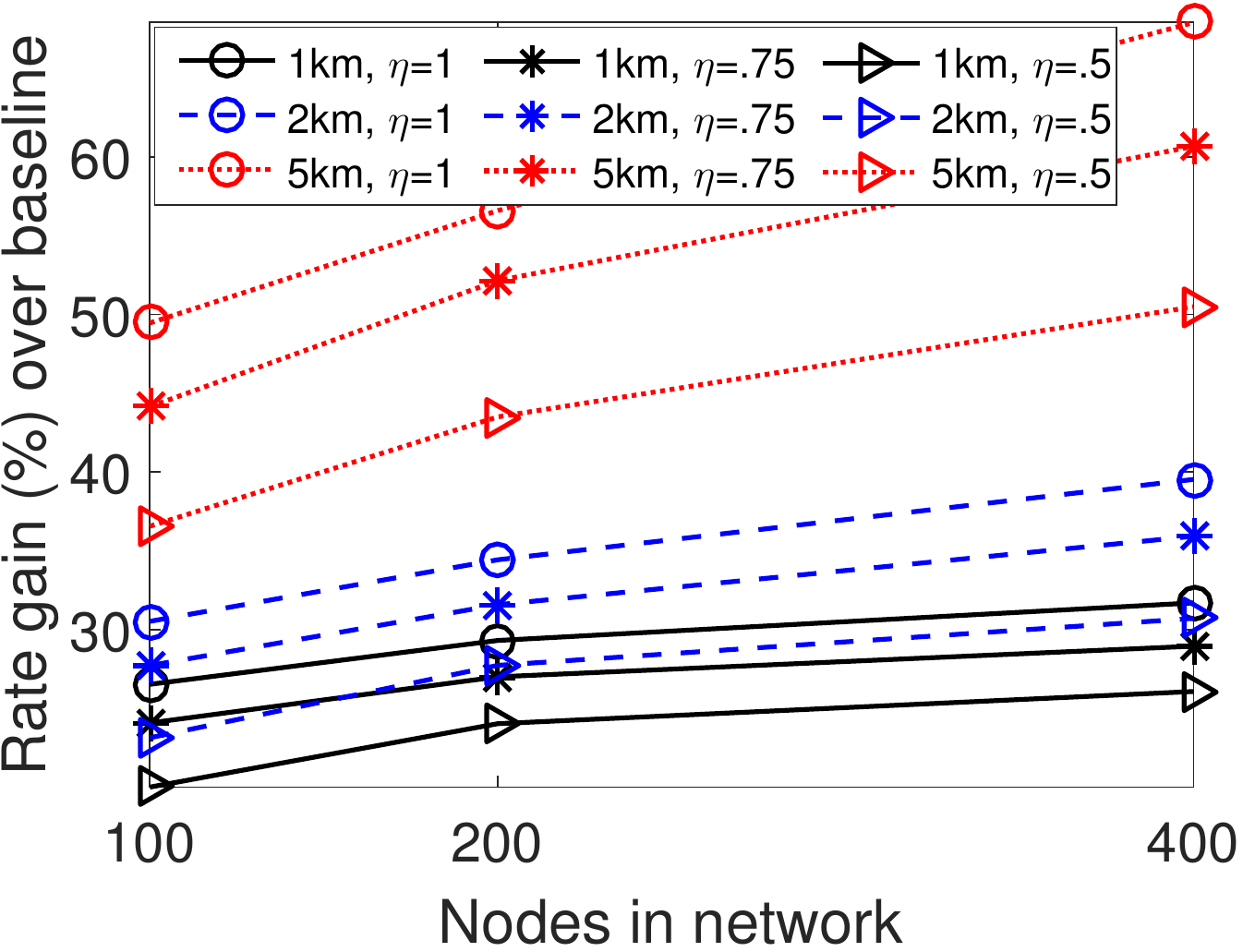}%
			\label{fig:evalGainsOverBaseline}
		}
		\subfloat[\small]{\includegraphics[width=.35\columnwidth]{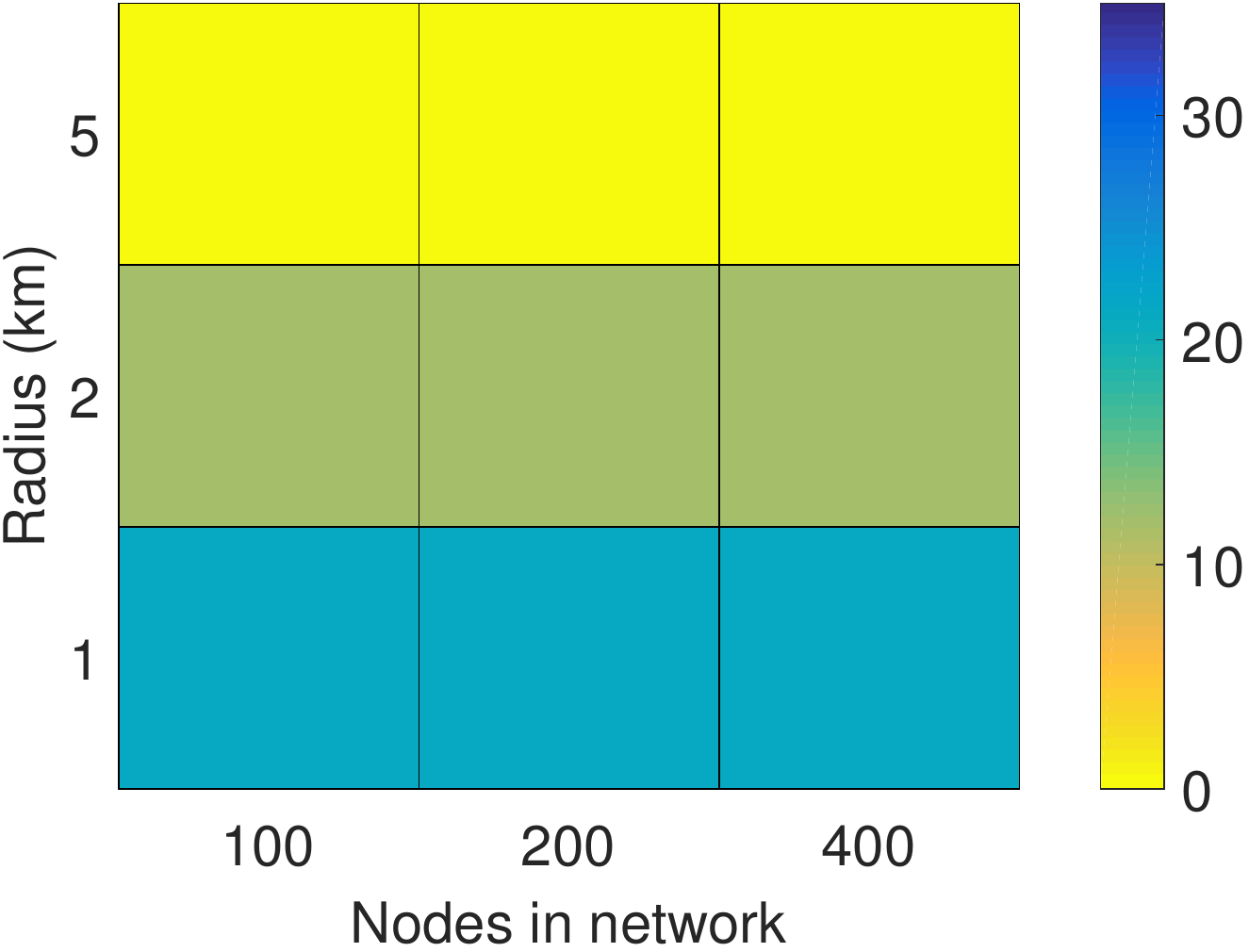}%
			\label{fig:pcolorAvgCellularSINR}
		}
		\subfloat[\small]{\includegraphics[width=.35\columnwidth]{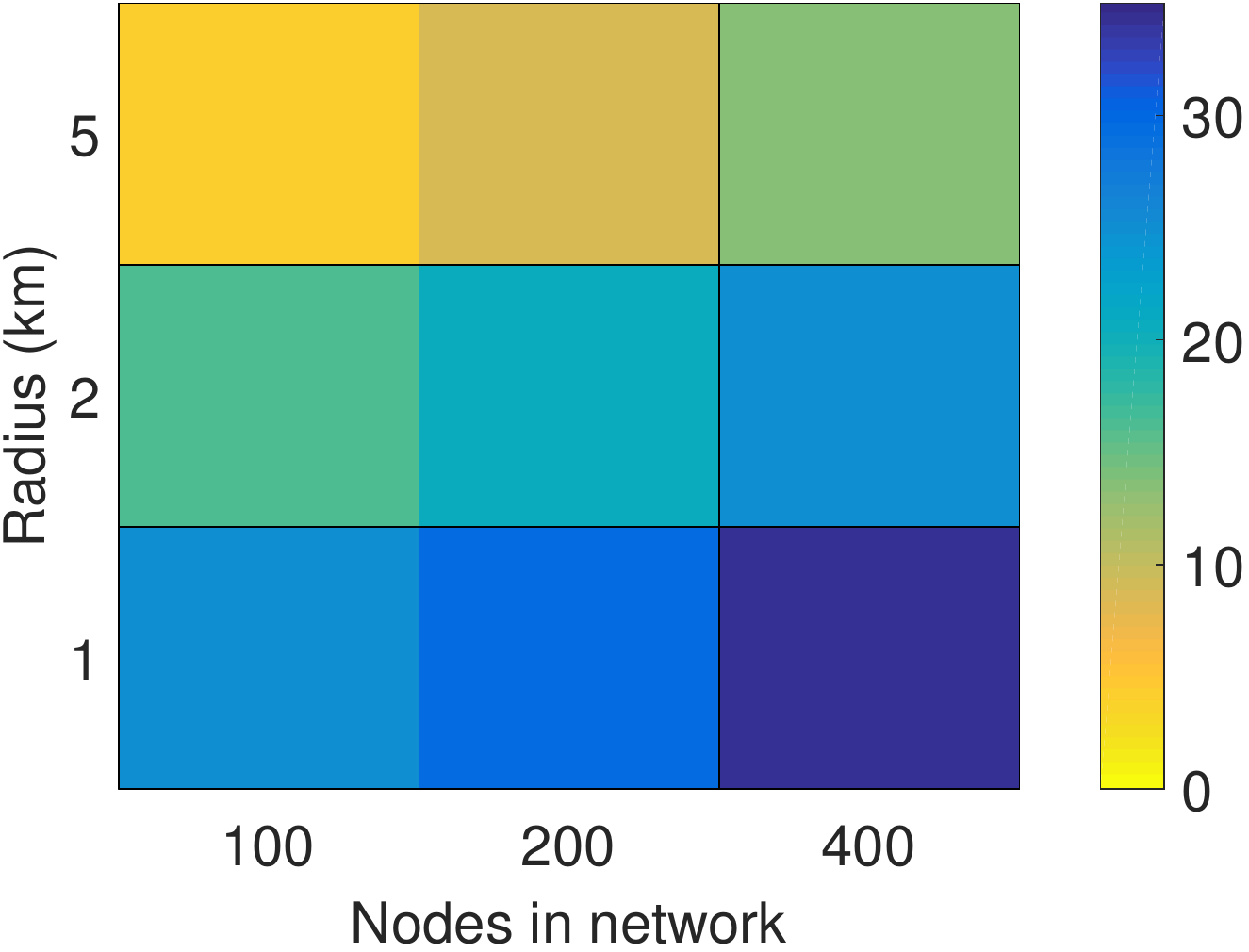}%
			\label{fig:pcolorAvgWiFiSINR}
		}
		\caption{\small (a) Gains in rates of nodes over baseline. (b) Average cellular SINR (dB). (c) Average largest WiFi SINR (dB) seen by a node. For all figures, both exponents $\alpha,\alpha_W = 3$.}			
	\end{center}
\label{fig:WiFiandCellularSINR}
	\vspace{-0.3in}
\end{figure}

We will use the metrics of (a) Shannon rate gain $\text{SRG}_j$, (b) Time share gain $\text{TSG}_j$, and (c) percentage gain $G_j$ in the rate of node $j$ over its baseline rate. They are defined as
\begin{align}
\text{SRG}_j = \frac{R_i^{(B)}}{R_j^{(B)}},\ 
\text{TSG}_j = \frac{(R_{j}/R_i)(1/H)}{(1/N)},\ G_j = \frac{R_j - R^{(B)}_j}{R^{(B)}_j} \times 100.
\label{eqn:evalMetrics}
\end{align}
$\text{SRG}_j$ is the ratio of the baseline rate of client $j$'s hotspot $i$ to its baseline rate. $\text{TSG}_j$ is the ratio of the time share of $j$ in the hotspot network to that in the baseline network. The time share of $j$ in the hotspot network is the product of the fraction $1/H$ of the total time that its hotspot's link gets to the tower and the fraction of time that $j$ gets on the hotspot's link, which is given by $R_{ij}/R_i = R_j/R_i$. For the rate $R_j$ of node $j$ in a hotspot network to be larger than its baseline rate $R^{(B)}_j$, the node must have either $\text{SRG}_j > 1$ or $\text{TSG}_j > 1$.

\emph{Gains in rate:} Figure~\ref{fig:evalGainsOverBaseline} shows the percentage gains $G_j$ in rate as a function of the number of nodes in the network, for different $\eta$ and $\mathcal{R}$. The exponents were set to $\alpha,\alpha_W = 3$. For a fixed radius (same line color and type) and number of nodes, gains increase as WiFi efficiency $\eta$ increases, because of larger WiFi link rates. Minimum rate gains of $20\%$ were obtained. 

\begin{figure}
	\begin{center}
	\subfloat[\small]{\includegraphics[width=.45\columnwidth]{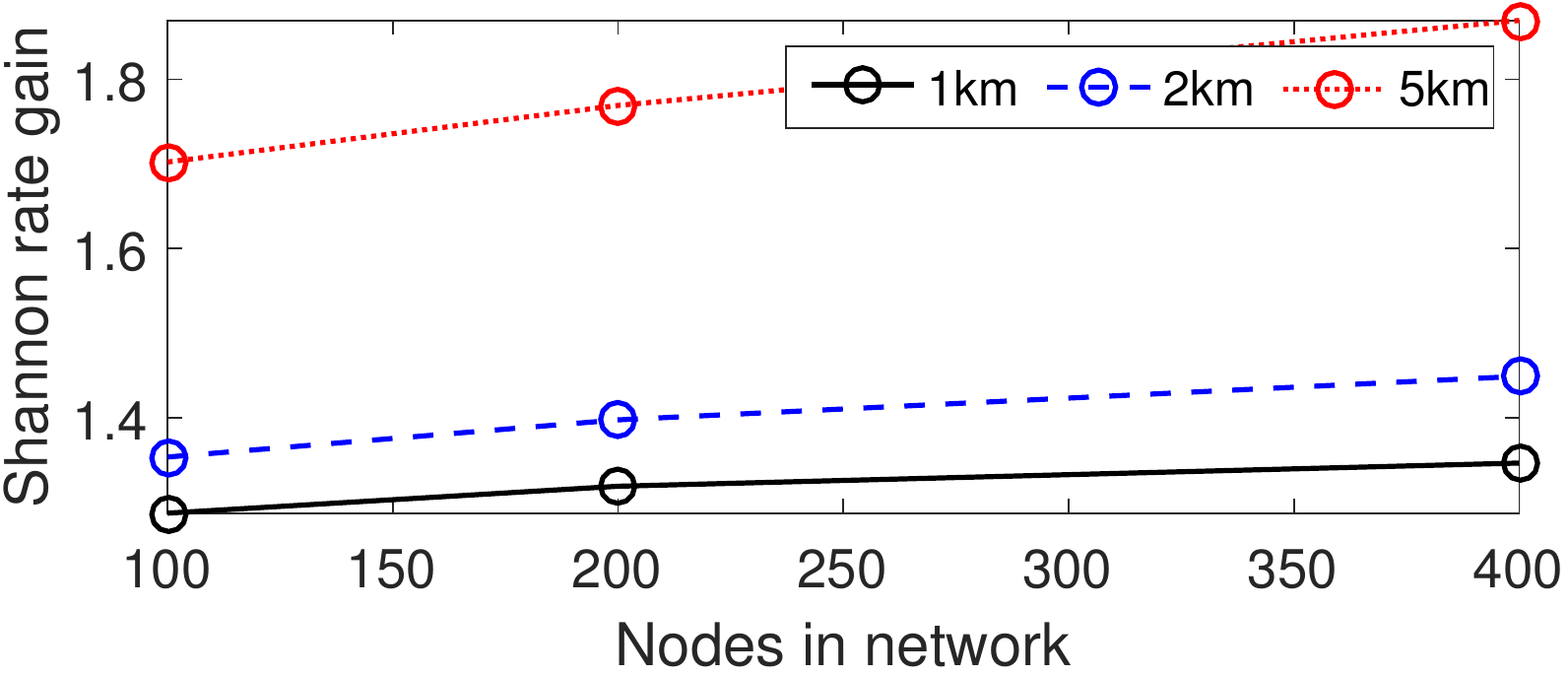}%
			\label{fig:evalAvgShannonRate}
		}\hspace{0.15in}
	\subfloat[\small]{\includegraphics[width=.45\columnwidth]{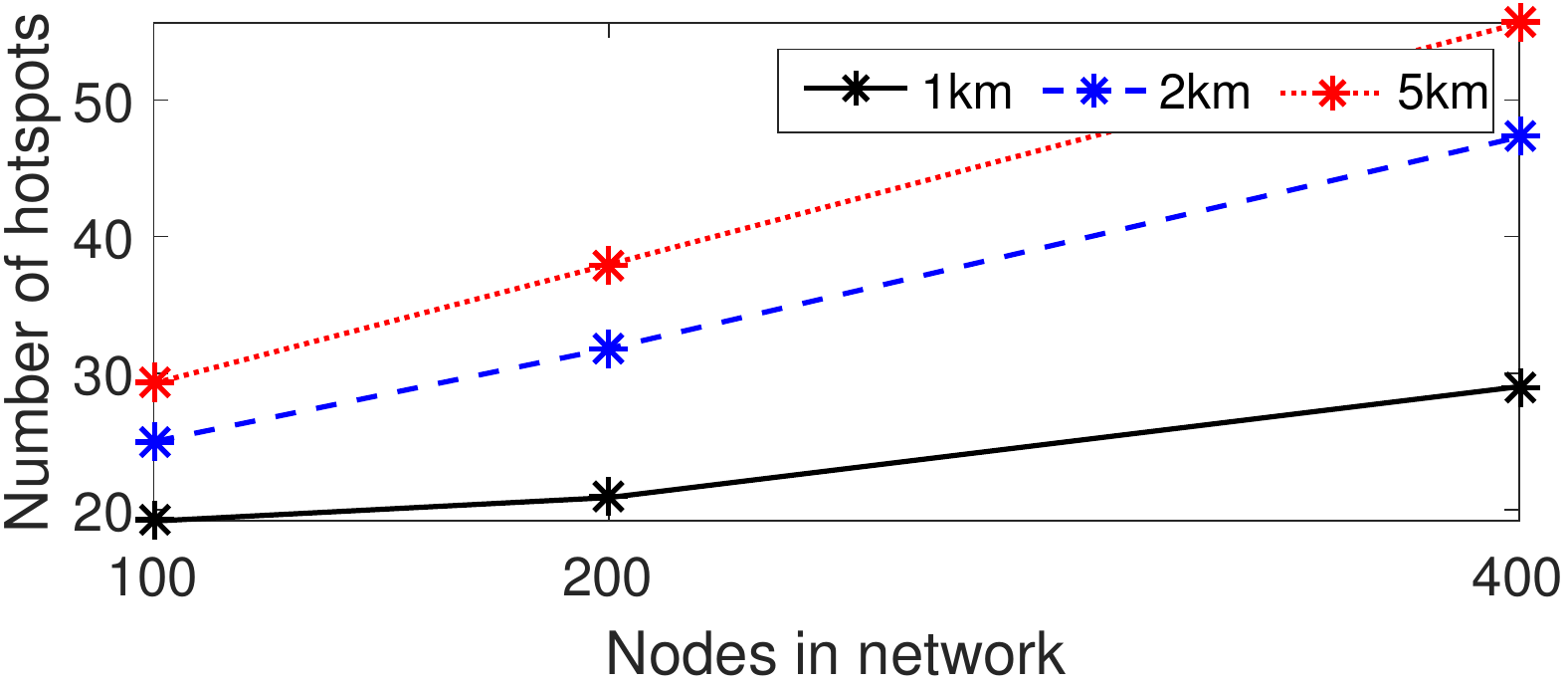}%
		\label{fig:evalNumHotspots}
	}
\caption{\small (a) SRG and (b) number of hotspots, as a function of number of nodes, for different radii and $\eta=0.75$.}			
	\end{center}
		\vspace{-0.3in}
\end{figure}

\emph{Fixing the number of nodes:} Consider an efficiency, say $\eta=0.75$, (marked by $*$) in Figure~\ref{fig:evalGainsOverBaseline}. For a fixed number of nodes, gains increase as radius increases. As radius increases, both WiFi (largest seen by a node) and cellular SINR(s) reduce. Figures~\ref{fig:pcolorAvgCellularSINR} and~\ref{fig:pcolorAvgWiFiSINR} summarize these changes. However, the WiFi SINR is about $5$ dB larger than the cellular SINR, even for the most sparse configuration of $N=100$ and $\mathcal{R}=5$ km. This allows for feasible hotspot networks. In addition, as shown in Figure~\ref{fig:evalAvgShannonRate}, as radius increases for a fixed number of nodes, the Shannon rate gains obtained by nodes in hotspot networks increase. This is because, networks with a large radius have a larger fraction of nodes with small SINR(s) (average $0$ dB for a $5$ km radius in Figure~\ref{fig:pcolorAvgCellularSINR}), which see large improvements in \emph{Shannon rate} due to the improvement in their SINR to tower, as a result of connecting via a hotspot. This explains the larger rate gains seen in Figure~\ref{fig:evalGainsOverBaseline} as radius increases for a fixed $N$.

\emph{Fixing the radius:} The gains in the baseline rate, shown in Figure~\ref{fig:evalGainsOverBaseline}, also increase as we increase the number of nodes $N$ for a fixed radius. While the average cellular SINR of a node is not affected as the number of nodes increases (see Figure~\ref{fig:pcolorAvgCellularSINR}), the baseline time share $1/N$ a node gets to the tower reduces. The increase in number of nodes, however, improves WiFi SINR (see Figure~\ref{fig:pcolorAvgWiFiSINR}). This is because, on an average, the distance of a node from another node decreases. Note that larger WiFi SINR is not a sufficient condition for a hotspot network to give rate gains over the baseline. It must be accompanied by a larger number of hotspots so that the number of clients per hotspot stays small as the number of nodes $N$ increases. 

In Figure~\ref{fig:evalNumHotspots} we see that the number of hotspots increase with the number of nodes, for a fixed radius. The number of hotspots, however, are much smaller than the total number of nodes. Together with large WiFi SINR this leads to large gains over baseline. Finally, note that as $N$ increases for a fixed radius, the Shannon rate gain (SRG), shown in Figure~\ref{fig:evalAvgShannonRate}, sees little change. This is because the cellular SINR(s) of the nodes stay similar as $N$ changes with radius fixed.

\begin{figure}
	\begin{center}
		\subfloat[\small]{\includegraphics[width=.45\columnwidth]{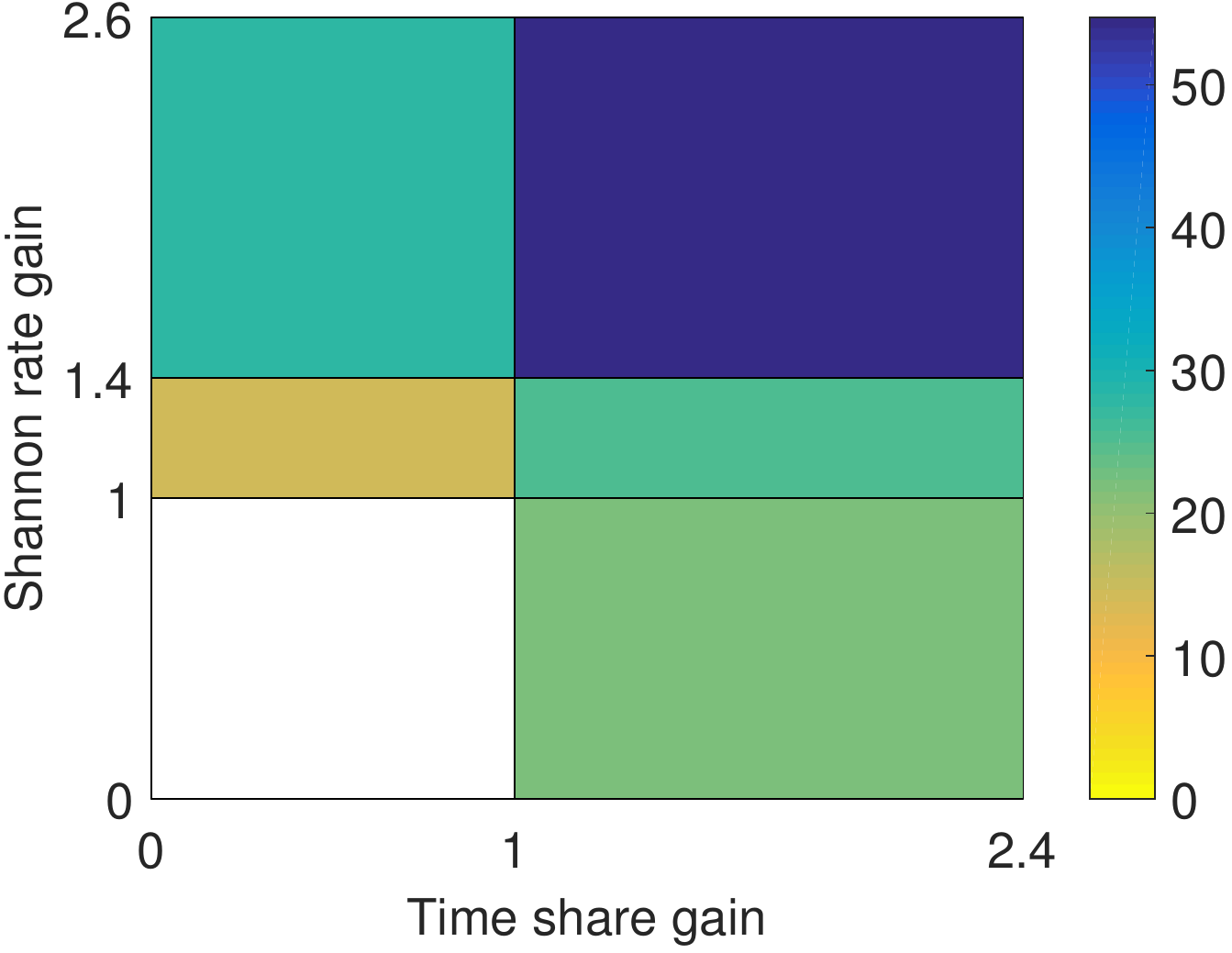}%
			\label{fig:pcolorAvgGain}
		}\hspace{0.5in}
		\subfloat[\small]{\includegraphics[width=.45\columnwidth]{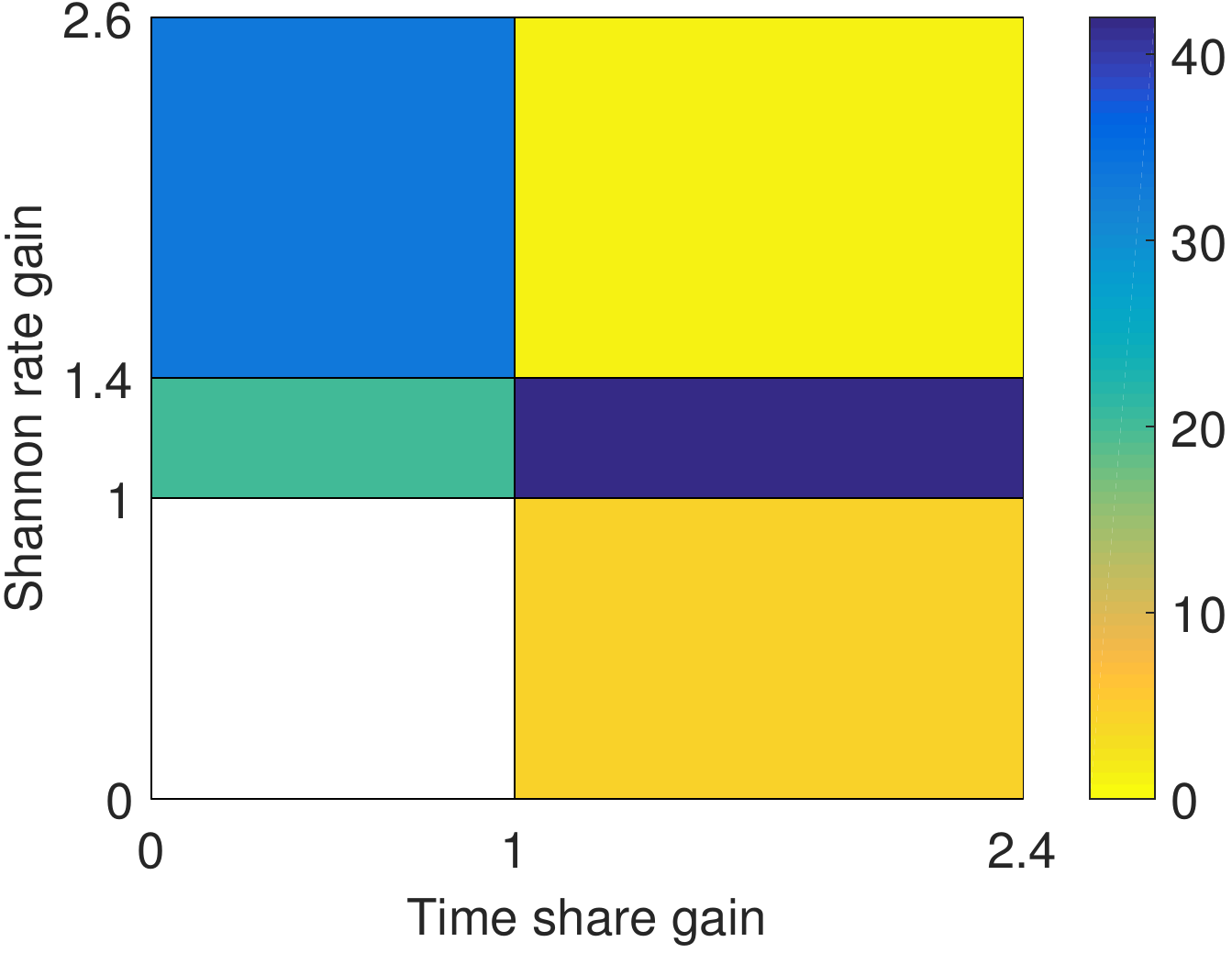}%
			\label{fig:pcolorNumNodes}
		}\hspace{0.1in}
		\caption{\small (a) Average rate gains of nodes in the network shown for different ranges of SRG and TSG. (b) Percentages of nodes that belong to different ranges. Shown for $N=100$, $\mathcal{R}=1$ km, $\eta=0.75$, and $\alpha,\alpha_W = 3$.}			
	\end{center}
		\vspace{-0.3in}
\end{figure}

Figures~\ref{fig:pcolorAvgGain} and~\ref{fig:pcolorNumNodes} explain rate gains obtained by nodes in terms of SRG and time share gain (TSG), for $N=100$, $\mathcal{R}=1$ km, and $\eta=0.75$. We quantize the SRG into the regions of $(0,1),\ (1,1.4),\ (1.4,2.6)$ and the TSG into the regions of $(0,1),\ (1,2.4)$. For each resulting region, Figure~\ref{fig:pcolorAvgGain} shows the rate gain obtained, on an average, by nodes whose SRG and TSG lie in the region. Figure~\ref{fig:pcolorNumNodes} shows the percentage of nodes that lie in each region.

\emph{Where do gains come from?:} From Figure~\ref{fig:pcolorNumNodes} we see that most nodes (about $96\%$) in the network see a SRG $\ge 1$. These include, (a) $20\%$ of all nodes that see a SRG in $(1,1.4)$ and a TSG in $(0,1)$, (b) a larger $28\%$ that see a SRG in $(1.4,2.6)$ and a TSG in $(0,1)$, and (c) $41\%$ that see a SRG in $(1,1.4)$ and a TSG in $(1,2.4)$. About $50\%$ of the $41\%$, which is $20\%$ of all nodes, are  hotspots. In fact, all hotspots lie in the SRG region $(1,1.4)$ and TSG region $(1,2.4)$. 

This implies that about $48\%$ of nodes (all clients, Shannon rate gain in $(1,2.6)$ and time share gain in $(0,1)$) see improvements in their rate by in effect trading their time share to the tower with their hotspot in exchange for a better SINR link to the tower. Amongst these clients, those in the Shannon rate region $(1,1.4)$ have average rate gains of $14\%$ and those in the region $(1.4,2.6)$ have gains of about $28\%$. The rate gains are shown in Figure~\ref{fig:pcolorAvgGain}.

The $41\%$ of nodes that see a SRG in $(1,1.4)$ and a TSG in $(1,2.4)$ obtain average gains of about $26\%$ (see Figure~\ref{fig:pcolorAvgGain}). Amongst these, the hotspots obtain average gains of $20\%$ and the clients that of $30\%$ (not shown in the figures). 

We are left with nodes that see a TSG in $(1,2.4)$ and a SRG either in $(0,1)$ ($4\%$ of nodes) or in $(1.4,2.6)$ ($1\%$ of nodes). The former are those that have a hotspot with a cellular SINR smaller than their own. The loss in SINR is, however, compensated by increase in time share. The latter are the very few that see large improvements both in Shannon rate and time share.

To summarize, all hotspots benefit from an increased time share. They don't see any benefits to their Shannon rate as their link to the tower is the same as in the baseline network. Most clients ($48\%$ of nodes) trade a smaller time share for a better link SINR. Fewer clients ($21\%$ of nodes) see reasonable improvements in both time share and link SINR. These summary observations hold for all networks we have simulated.
\begin{figure}[tb]
\begin{center}
\subfloat[\small]{
 \includegraphics[width=.45\columnwidth]{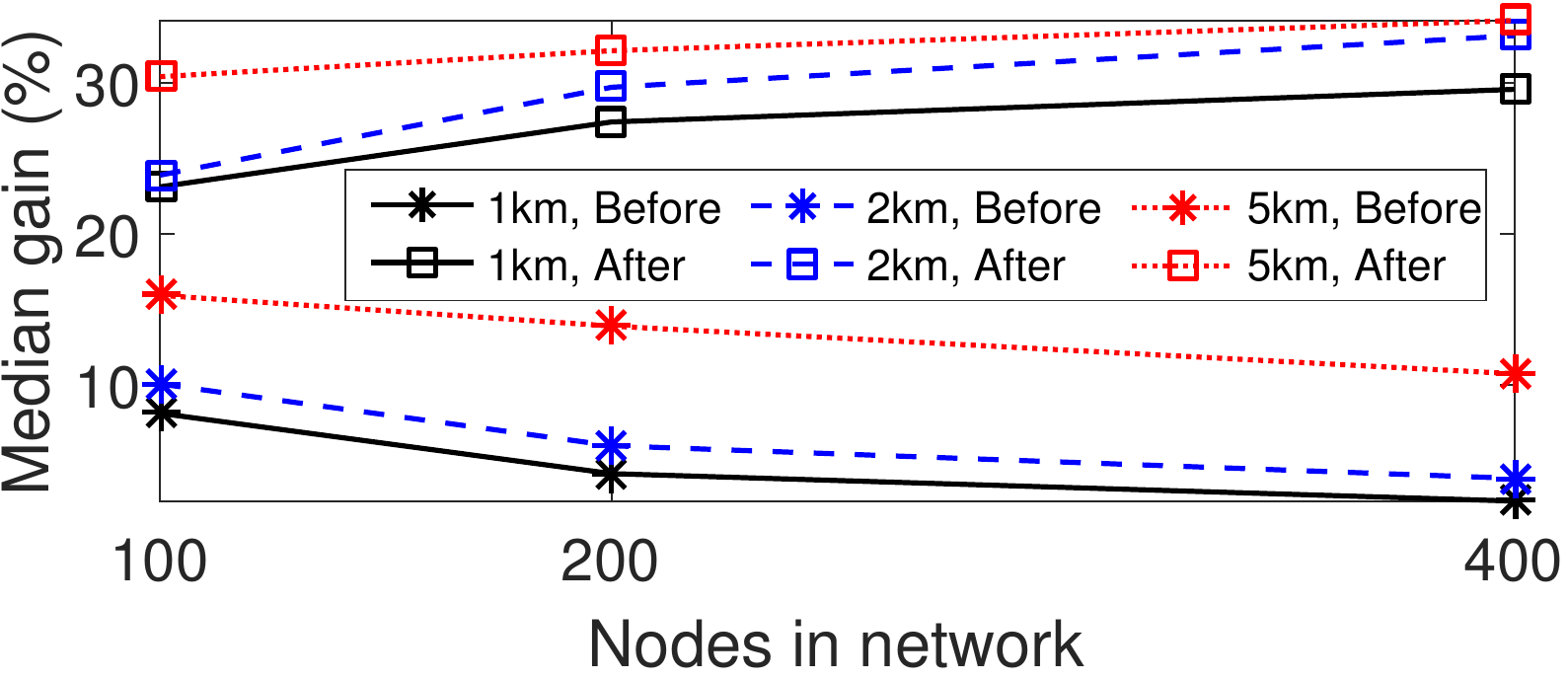}%
\label{fig:evalmedianBeforeAfterLoading}%
}\hspace{0.1in}
\subfloat[\small]{
\includegraphics[width=.45\columnwidth]{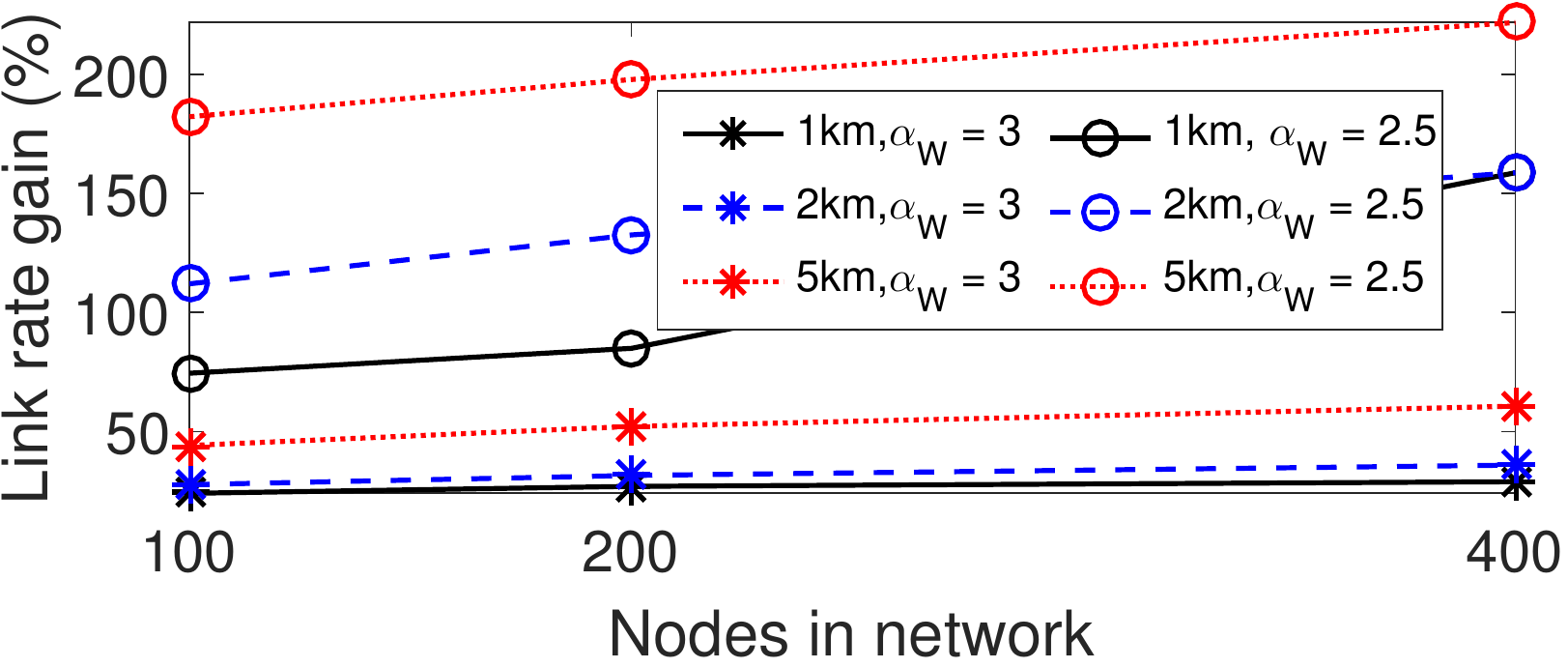}%
\label{fig:evalGainsAllExponentOverBaseline}
}
\caption{(a) Median rate gains before and after application of Algorithm~\ref{alg:fairLoading} for $\eta=0.75$, and $\alpha,\alpha_W = 3$. (b) Rate gains compared for when $\alpha_W=3$ and $\alpha_W=2.5$. We have $\eta=0.75$ and $\alpha=3$.}
\end{center}
\vspace{-0.3in}
\end{figure}

We end this section with Figures~\ref{fig:evalmedianBeforeAfterLoading} and~\ref{fig:evalGainsAllExponentOverBaseline}. Figure~\ref{fig:evalmedianBeforeAfterLoading} shows the significant improvements in \emph{median} rate gain due to the fair loading that results from Algorithm~\ref{alg:fairLoading} (Compute-Fair-Loading). Larger median rates imply that the gains in sum rate of the network are shared more fairly among nodes in the hotspot network. Figure~\ref{fig:evalGainsAllExponentOverBaseline} compares the case of $\alpha_W=3$ and $\alpha_W=2.5$. Note that setting $\alpha_W=2.5$ when $\alpha = 3$ makes the WiFi channel a lot better than the cellular channel. The resulting gains, as seen in the figure, are much larger when $\alpha_W=2.5$.

\section{Conclusions}
\label{sec:conclusions}
We formulated the sum rate optimization problem that splits nodes connected to the cellular network into nodes who will be configured as hotspots and nodes who will access the Internet by becoming clients of the hotspots. In the hotspot network, all nodes must at least get the rate they were getting when each one of them was directly connected to the cellular network. We proposed a heuristic approach to solve the problem, which is a MINLP, and evaluated its efficacy via extensive simulation.

\bibliographystyle{IEEEtranTCOM}
\bibliography{IEEE_TCOM}

%




\end{document}